\documentclass[prd,twocolumn,floatfix,showpacs,preprintnumbers,nofootinbib]{revtex4}
\usepackage[utf8x]{inputenc}
\usepackage{mathrsfs}
\usepackage{amssymb}
\usepackage{amsmath}
\usepackage{graphicx}

\usepackage[caption = false]{subfig}
\usepackage[normalem]{ulem}
\usepackage{xcolor}
\definecolor{lcolor}{rgb}{0.5,0,0}
\definecolor{citcolor}{rgb}{0,0.3,0.0}

\usepackage[breaklinks,colorlinks,urlcolor=blue,citecolor=citcolor,linkcolor=lcolor]{hyperref}
\usepackage{mciteplus}

\newcommand{\rt}{{\mathbf{r}}}

\newcommand{\bt}{{\mathbf{b}}}

\newcommand{\Deltat}{{\boldsymbol{\Delta}}}

\newcommand{\ud}{\, \mathrm{d}}

\newcommand{\nc}{{N_\mathrm{c}}}

\newcommand{\gev}{\ \textrm{GeV}}
\newcommand{\fm}{\ \textrm{fm}}

\newcommand{\lqcd}{\Lambda_{\mathrm{QCD}}}
\newcommand{\as}{\alpha_{\mathrm{s}}}
\newcommand{\aem}{\alpha_{\mathrm{em}}}

\newcommand{\sigmaa}{{ \sigma^A_\textrm{dip} }}

\newcommand{\sigmadip}{{ \sigma_\textrm{dip} }}

\newcommand{\xpom}{{x_\mathbb{P}}}

\newcommand{\der}{\mathrm{d}}

\newcommand{\A}{{\mathcal{A}}}

\begin{document}

\author{H. M\"antysaari}
\affiliation{Department of Physics, University of Jyväskylä,
 P.O. Box 35, 40014 University of Jyväskylä, Finland}

\author{P. Zurita}
\affiliation{
Physics Department, Brookhaven National Laboratory, Upton, NY 11973, USA
}

\title{
In depth analysis of the combined HERA data in the dipole models with and without saturation
}

\pacs{13.60.-r,24.85.+p}
%
\preprint{}

\begin{abstract}
We present an updated impact parameter dependent saturation model (IPsat) determined trough a fit to the combined HERA I and I+II reduced cross section data. The same HERA data are used to fit the linearized (IPnonsat) version of the applied dipole amplitude, which makes it possible to estimate the magnitude of the saturation effects in various experiments. We find that both parametrizations provide comparable descriptions of the considered data when an effective confinement scale dynamics is incorporated with quark masses. Moreover, it is possible to consistently determine the light and charm quark masses. The role of potentially non-perturbatively large dipoles is examined in detail, with the result that, especially in case of the structure function $F_2$, their contribution is numerically significant. Potential to discriminate between the two models in future $e+p$ and $e+A$ experiments is also illustrated.
\end{abstract}

\maketitle

\section{Introduction}

Quantum Chromodynamics, the theory of the strong interaction, is a vast field with a plethora of diverse phenomena still unexplored. Despite being the object of study of the high energy and nuclear physics communities both theoretically and experimentally, the true nature of a proton, its constituents, their interactions and how they come together to conform it remain elusive. In the collinear framework at a given resolution scale $Q^{2}$, the proton can be described as composed of quarks and gluons carrying a fraction $x$ of the proton momentum. Once known at an initial scale $Q^{2}_{0}$, the partonic densities can be determined at any scale $Q^{2}$ using the DGLAP evolution equations \cite{Gribov:1972ri,Gribov:1972rt,Altarelli:1977zs,Dokshitzer:1977sg}. This picture is successfully supported by extensive experimental evidence; however it can not be valid for all the kinematic range: as one explores lower $x$ values the DGLAP equations predict an infinite rise of the gluon density which would break unitarity. It follows that some phenomena, e.g. gluon recombination, has to enter in order to tame this dangerous behavior. This dynamically generates the \emph{saturation scale} $Q_s^2$, which determines when the transition to the non-linear regime of QCD takes place. Moreover, if $Q_s^2$ is large, perturbative calculations become possible as the strong coupling constant is weak. A successful theoretical framework to describe QCD in this region is known as the Color Glass Condensate~\cite{Gelis:2010nm}. 

There are many theoretical models that incorporate saturation to QCD calculations with different approaches and considerations, a popular one being the IPsat parametrization \cite{Kowalski:2003hm,Kowalski:2006hc,Rezaeian:2012ji,Luszczak:2013rxa,Luszczak:2016bxd}. The parameters that provide the necessary non-perturbative input to these models are determined through fits to available data, the bulk of them being high precision deep inelastic scattering (DIS) data measured at HERA in electron-proton ($e+p$) collisions \cite{Aaron:2009aa,Abramowicz:1900rp,Abramowicz:2015mha,H1:2018flt}. Despite the wide kinematic range covered by this collider there are no spectacular signals of deviation from the DGLAP predictions and some observed discrepancies might be due to reasons other than saturation. Recently a possible hint of a non-linear regime from HERA data at low $x$ was shown to be 
feasibly explainable
by the inclusion of resumed logarithmic corrections \cite{Ball:2017otu}. Saturation model calculations have also been able to provide a natural description for the nearly flat center-of-mass energy dependence of the diffractive to total cross section ratio at HERA~\cite{GolecBiernat:1998js,GolecBiernat:1999qd} (see also Ref.~\cite{Kowalski:2008sa}).
In general, there is no clear consensus on whether or not the onset of saturation has been reached and it will be necessary to perform a thorough and detailed exploration of the kinematic space beyond our current knowledge in order to observe the non-linear regime of QCD. Future facilities such as the Electro-Ion Collider (EIC) \cite{Accardi:2012qut,Aschenauer:2017jsk}, the Large electron-Hadron Collider (LHeC) \cite{AbelleiraFernandez:2012cc} and the Future Circular Collider (FCC-eh)~\cite{Zimmermann:2014qxa} hold the key to this door.        

In this study we present a new determination of the IPsat and its linearized version (``IPnonsat'') description of the HERA combined data \cite{Aaron:2009aa,Abramowicz:1900rp,Abramowicz:2015mha,H1:2018flt} in the framework of the dipole model. What is new here, compared to the previous literature~\cite{Kowalski:2003hm,Kowalski:2006hc,Rezaeian:2012ji}, is that we also fit the IPnonsat model parametrization to the precise combined HERA data which allows us to explore the expected magnitude of saturation effects in current and future collider experiments. In addition, by simultaneously fitting the total cross section and the charm contribution to it, it becomes possible to determine the quark masses in this framework. For consistency, a variable flavor number scheme is also applied.

This work is organized as follows. In Sec. \ref{dipole} we describe the inclusive photon-proton interaction in terms of the dipole model for both IPsat and IPnonsat formulations. The analysis of the combined inclusive and charm data from HERA is the subject of Sec. \ref{HERA}, while in Sec. \ref{amplitude} we present an analysis of the obtained dipole-proton scattering amplitude. The application of our determined parameters to the exclusive production of vector mesons is discussed in Sec. \ref{EVM}. In Sec. \ref{Future} we consider the potential of the EIC and the LHeC to provide a signal of saturation in both $e+p$ and $e+A$ collisions. Finally Sec. \ref{conclusions} summarizes our findings.

\section{Photon-proton scattering in the dipole picture}
\label{dipole}
The most precise study of the proton structure has been performed in 
$e+p$ DIS experiments at HERA~\cite{Aaron:2009aa,Abramowicz:2015mha,Abramowicz:1900rp,H1:2018flt}. The process is described as the electron emitting a virtual photon with momentum $q$, which then probes the proton with a resolution scale $Q^{2}=-q^{2}$. In the dipole picture, applicable at high energy and not too large $Q^2$, the virtual photon-proton scattering process can be factorized in two parts: the $\gamma^{*}$ splitting into a $q\bar{q}$ pair, and the dipole-target scattering. The total $\gamma^* p$ cross section is subsequently obtained as the imaginary part of the forward elastic $\gamma^* p \to \gamma^* p$ scattering amplitude using the optical theorem. The photon splitting into a dipole with transverse separation $\rt$ is described in terms of the photon wave function $\Psi^f_{L,T}(r,z,Q^2)$, where $f$ is the quark flavor, $L$ and $T$ refer to transverse and longitudinal polarizations, $|\rt|=r$, and $z$ is the fraction of the longitudinal momentum of the photon carried by the quark. The total photon-proton cross section is then given by
\begin{equation}
\sigma^{\gamma^* p}_{L,T}(x,Q^2) = \int \der^2 \bt \der^2 \rt \int_0^1 \frac{\der z}{4\pi} |\Psi_{L,T}^f(r,z,Q^2)|^2 \frac {\ud \sigmadip}{\ud^2 \bt}, 
\end{equation}
where $\frac {\ud \sigmadip}{\ud^2 \bt}$ is the proton-dipole cross-section with $\bt$ denoting the impact parameter, and one has to sum over all the quark flavors included in the analysis ($u,d,s,c$ and $b$ in this work). The photon wave functions for the transverse and longitudinal polarizations summed over spins and helicities read~\cite{Kovchegov:2012mbw}
\begin{multline}
\label{eq:photon_t}
	|\Psi_T(r,z,Q^2)|^2 = \frac{2\nc}{\pi} \aem e_q^2 \Big\{ [ z^2+(1-z)^2 \varepsilon^2 K_1^2(\varepsilon r) \\
	+ m_f^2 K_0^2(\varepsilon r) ] \Big\}
\end{multline}
and
\begin{equation}
\label{eq:photon_l}
	|\Psi_L(r,z,Q^2)|^2 = \frac{8 \nc}{\pi} \aem e_q^2 Q^2 z^2(1-z)^2 K_0^2(\varepsilon r),
\end{equation}
with $\varepsilon^2 = z(1-z)Q^2 + m_q^2$. Here, $e_q$ is the fractional charge of the quark $q$ and $m_q$ is the quark mass.

The proton structure functions $F_2$ and $F_L$ can be written in terms of the total photon-proton cross section as
\begin{align}
	F_2 &= \frac{Q^2}{4\pi^2 \aem} (\sigma^{\gamma^*p}_L + \sigma^{\gamma^*p}_T), \\
	F_L &= \frac{Q^2}{4\pi^2 \aem} \sigma^{\gamma^*p}_L.
\end{align}
The most precise combined HERA data is released in the form of the reduced cross section
\begin{equation}
	\sigma_r(x,y,Q^2) = F_2(x,Q^2) - \frac{y^2}{1+(1-y)^2} F_L(x,Q^2),
\end{equation}
where $y=Q^2/(xs)$ is the inelasticity of the $e+p$ scattering with center-of-mass energy $\sqrt{s}$.

In the IPsat model the saturation scale of the target depends on the impact parameter, and the cross section is written as
\begin{equation}
\label{eq:ipsat}
\frac {\ud \sigmadip}{\ud^2 \bt} = 2\left[1 - \exp \left( -r^2 F(x,r) T_p(\bt) \right) \right].
\end{equation}
The proton density profile $T(\bt) = e^{-\bt^2/(2B_p)} / (2\pi B_p)$ is assumed to be Gaussian, and we use fixed $B_p=4\gev^{-2}$ based on HERA exclusive $J/\Psi$ production data. Thus, the effective transverse area of the proton is not a free parameter in the model, and the root mean square radius of the proton is $\sqrt{2 B_p}$. However, we note that this parametrization describes only part of the observed growth of the proton already at the HERA energies~\cite{Aktas:2005xu,Chekanov:2002xi} due to the Gribov diffusion. Including this effect would require us to either parametrize the proton width $B_p$ and try to fit it simultaneously to the HERA data, or solving impact parameter dependent small-$x$ evolution equations such as in Refs.~\cite{Berger:2011ew,Berger:2012wx,Schlichting:2014ipa}. We do not want to include exclusive data into our fit due to additional model uncertainties, and we leave it for a future work.

 At the lowest order in perturbation theory the function $F$ is proportional to the DGLAP evolved gluon distribution function
\begin{equation}
F(x, r^2) = \frac{\pi^2}{2 \nc} \as(\mu^2) xg(x, \mu^2),
\end{equation}
with $x$ being the momentum fraction of the proton carried by the gluon, and the scale $\mu^2$ is a function of $r^2$. This parametrization gives the correct pQCD limit for the dipole cross section, $\sigmadip \sim r^2$. At large dipoles, the saturation effects are described by having an eikonalized gluon distribution function, which gives $\ud \sigmadip / \ud^2 \bt \to 2$ at large $r$, corresponding with the interaction probability being unity at large dipoles.

The scale at which the gluon density and the strong coupling constant are evaluated is chosen to be $\mu^2 = \mu_0^2 + C/r^2$. Here, unlike in previous fits~\cite{Kowalski:2006hc,Rezaeian:2012ji}, we fix $\mu_{0}^{2}=1.1\gev^2$ and let $C$ be a free parameter. This allows us to force $\mu^2$ to remain in the perturbative region. In our fit we only include data in the $Q^2$ bins that satisfy $Q^2>\mu_0^2$, which guarantees the applicability of the perturbative calculation. We also consider data in the kinematical region $x<0.01$ where the dipole picture can be considered to be most reliable.

The gluon density at the initial scale $\mu_0$ is parametrized as
\begin{equation}
\label{eq:xg_ic}
xg(x,\mu_0^2) = A_g x^{-\lambda_g} (1-x)^{6},
\end{equation}
where $A_g$ and $\lambda_g$ are free parameters to be determined by the fit. Unlike previous works, we use a variable flavor number scheme when evaluating the strong coupling constant $\as$ and when solving the DGLAP evolution for the gluon distribution. Neglecting the change in the number of flavors and using a fixed number of quark flavors as the data moves in $Q^{2}$ is not the most adequate strategy from the theoretical point of view but, in practice, it is solely reflected in different values of the fitted parameters, without a sizable effect in the description of the data in the currently probed kinematic range. 

For simplicity we refrain from including the quark singlet contribution to the DGLAP evolution which should also be present. However we have checked that the fit quality and the resulting dipole amplitude are not significantly affected by its inclusion, and that the fit drives the quark singlet contribution to zero at the initial scale. Furthermore we choose the high-$x$ behavior to be an integer exponent ($6$) instead of the standard $5.6$ in order to speed up the DGLAP evolution performed in Mellin space. We have checked that this exponent does not have a significant impact on the determination of the parameters. The strong coupling constant is required to satisfy $\as(M_z = 91.1876\gev) = 0.1183$~\cite{Abramowicz:2015mha}. When evaluating the heavy quark contribution, the Bjorken $x$ is replaced by
\begin{equation}
\label{eq:xshift}
x_q= x\left(1 + \frac{4 m_q^2}{Q^2} \right)
\end{equation}
in order to take into account the kinematical shift, where $q$ refers to the quark flavor $c$ or $b$. 
As we stay in the perturbative (large $Q^2$) region in this work, the shift \eqref{eq:xshift} would have negligible effect in case of light quarks.
The quark masses $m_f$ for the light and charm quark are kept as free parameters and constrained by the fit.  In the fit we only include data that satisfy $x_c<0.01$ for the charm quark. The $b$ quark mass is set to $4.75\gev$, and the $b$ quark contribution to the structure function is included if the corresponding Bjorken-$x$ satisfies $x_b<0.1$. 

Effectively in the IPsat model we fit the $x$ dependence of the cross section to the HERA data, and extrapolate it down to smaller values of Bjorken-$x$ when calculating predictions e.g. for the future DIS experiments. The other approach used in small-$x$ phenomenology is to evolve the dipole amplitude in $x$ using the perturbatively derived Balitsky-Kovchegov evolution equation~\cite{Kovchegov:1999yj,Balitsky:1995ub}, and incorporate some of the DGLAP effects in the initial condition of the evolution, which is fitted to the HERA data. Currently these fits are done at the leading logarithmic accuracy with running coupling corrections, and very good description of the HERA data is obtained if the $x$ evolution speed is also adjusted in the fit process by fitting the coordinate space scale at which the running coupling constant is evaluated~\cite{Albacete:2010sy,Lappi:2013zma} (note that our fit parameter $C$ has a similar effect by controlling the scale at which we evaluate $\as(\mu^2)$ and $xg(x,\mu^2)$). Recently, there has been a lot of progress in developing the theory to NLO accuracy, see e.g. Refs~\cite{Balitsky:2008zza,Lappi:2015fma,Lappi:2016fmu,Lappi:2016oup,Boussarie:2016bkq,Beuf:2017bpd,Ducloue:2017ftk,Hanninen:2017ddy}.

In order to quantify the magnitude of the saturation effects, we also study the linearized version of the IPsat parametrization~\eqref{eq:ipsat}:
\begin{equation}
\label{eq:ipnonsat}
\frac {\ud \sigmadip}{\ud^2 \bt} = 2 r^2 F(x,r) T_p(\bt),
\end{equation}
to which we refer as the \emph{IPnonsat} model.
We emphasize that the rigorous way to look for the saturation effects is to compare saturation model calculations with the perturbative QCD results obtained by applying collinear factorization. In practice, however, comparing IPsat and IPnonsat results can be used as a first estimate for the strength of the saturation effects in the given process. In order to enable such a comparison, we fit the IPnonsat model parameters to the same HERA data.

\section{Description of the HERA reduced cross section data}
\label{HERA}

\begin{table*}[htbp]
  \centering
  \begin{tabular}{@{} cccccccccccc @{}}
    \toprule
    Type & HERA  & $\chi^2/N$ & $N$ & $Q^2_\text{min}$ & $Q^2_\text{max}$ & $m_l$ & $m_c$   & $C$ & $A_g$ & $\lambda_g$   \\
    \hline
    IPsat & I  & 1.0978 & $156+33$ & 1.5 & 50 & 0.03 & 1.3528 & 2.2894 & 2.1953 & 0.08289  \\ 
    IPsat & I+II & 1.2781 & $410+33$ & 1.5 & 50 & 0.03 & 1.3210 & 1.8178 & 2.0670 & 0.09575\\
    IPsat & I & 1.2634 & $229+42$ & 1.5 & 500 & 0.03 & 1.3296 & 2.6477 & 2.2097 & 0.07795 \\ 
  IPsat & I+II & 1.3014 & $609+42$ & 1.5 & 500 & 0.03 & 1.3113  & 2.3700 & 2.1394  & 0.08388 \\
    \hline
    IPnonsat & I & 1.122 & $156+33$ & 1.5 & 50 & 0.1516 & 1.3504 & 4.2974 & 3.0391 & -0.006657 \\ 
    IPnonsat & I+II & 1.3023 & $410+33$ & 1.5 & 50  & 0.1497 & 1.3180 & 3.5445 & 2.8460 & 0.008336    \\
         IPnonsat  & I & 1.2194 & $229+42$  & 1.5 & 500 & 0.1332 & 1.3187 & 5.6510 & 3.2820 & -0.03460 \\ 
         IPnonsat & I+II & 1.2944 & $609 + 42$& 1.5 & 500 & 0.1359 & 1.3047 & 4.7328 & 3.0573 & -0.01656
   
  \end{tabular}
  \caption{All dimensionfull parameters in $\gev$. $X+Y$ points means $X$ points for $\sigma_r$ and $Y$ points for $\sigma_{r,\text{charm}}$. Bottom mass is $m_b=4.75\,\gev$, and $B_p=4\gev^{-2}$. In the IPsat fit the light quark mass is fixed to prevent numerical instability. The starting scale for the DGLAP evolution is $\mu_0^2=1.1\gev^2$. Fit results with HERA I data~\cite{Aaron:2009aa,Abramowicz:1900rp} and HERA I+II data~\cite{Abramowicz:2015mha,H1:2018flt} are shown separately. }
    \label{tab:fits}
\end{table*}

The H1 and ZEUS experiments from HERA have published two combined datasets for the reduced cross section: the first one in Ref.~\cite{Aaron:2009aa} with the charm contribution in Ref.~\cite{Abramowicz:1900rp} where the HERA-I data are combined, and the latest final combined result for the inclusive reduced cross section including all HERA (HERA I+II) data in Ref.~\cite{Abramowicz:2015mha}. Recently the latest charm reduced cross section data from HERA I+II have been made available~\cite{H1:2018flt}. We will perform fits to both HERA I and HERA I+II datasets, but we will consider the fit to HERA I data to be our main result, as the charm cross section from HERA I+II is not yet published and the additional data in the newer dataset mostly affects the high-$Q^2$ region not included in the analysis. Moreover, the newer total reduced cross section dataset has more than twice as many data points in the region of interest of this work, which renders that data less sensitive to the charm quark if one does not introduce artificial weight factors (the HERA I+II charm reduced cross section data has as many points as in the HERA I results).

We include data in the region $1.5 < Q^2 < 50 \gev^2$. The lower limit, which we require to be larger than $\mu_0^2$, guarantees that there is a large scale justifying the perturbative calculation. As the validity of the dipole picture becomes questionable at high $Q^2$, we only include data up to $Q^2=50\gev^2$ in our main fit, though we also show results for fits done in the larger virtuality range with $Q^2_\text{max}=500\gev^2$. 

The free parameters in this work are $A_{g}$ and $\lambda_g$ that describe the gluon distribution at $\mu^2=\mu_0^2$, and $C$ that controls the momentum space scale corresponding to the given dipole size $|\rt|$. In addition, and as mentioned in the previous section, the light quark and charm quark masses are to be determined by the fit. However, as the bottom quark contribution is small, the fit can not reliably determine the $b$ quark mass and we set it to $m_b=4.75\gev$.

\begin{figure}[tb]
\begin{center}
\includegraphics[width=0.49\textwidth]{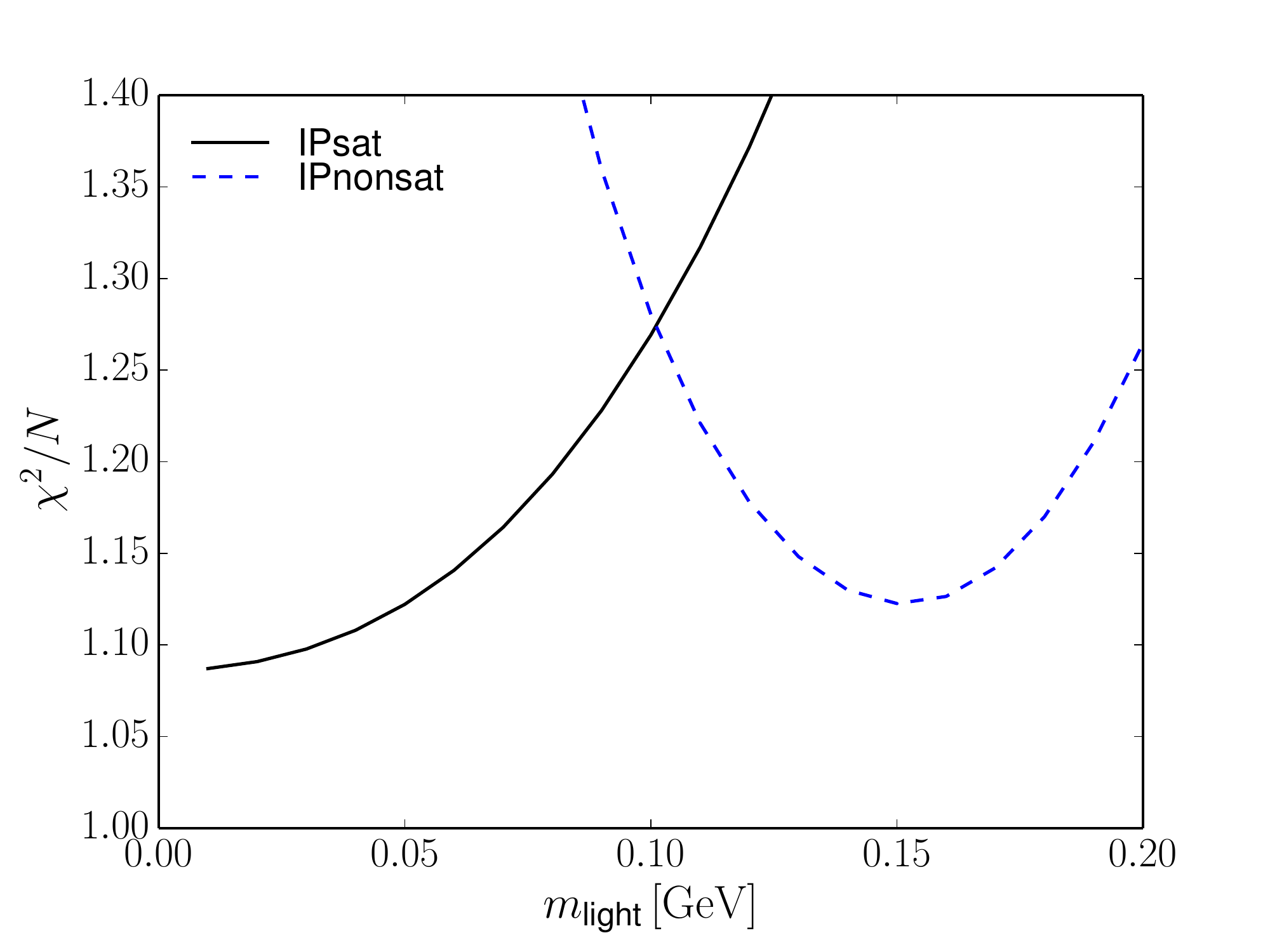}
\end{center}
\caption{
Fit quality to HERA inclusive~\cite{Aaron:2009aa} and charm~\cite{Abramowicz:1900rp} reduced cross section data as a function of the light quark mass.
}\label{fig:light_mass}
\end{figure}

The dependence of the fit quality on the light quark mass $m_\text{light}$ is shown in Fig.~\ref{fig:light_mass} when we fit the HERA I data~\cite{Aaron:2009aa,Abramowicz:1900rp}. Throughout this work we show results obtained by using the fit to the HERA I data. Here, the charm mass and all the other parameters are allowed to vary freely with the fixed light quark mass. As one moves to lower values of $m_\text{light}$ in the IPsat fits, the $\chi^2$ reaches a plateau, making it hard to determine a best fit extraction of it's value, similarly as in Ref.~\cite{Rezaeian:2012ji}. Therefore, and in order to have finite quark mass to act as an infrared regulator, we fix $m_\text{light}=0.03 \gev$ for the IPsat case. The situation is different in the IPnonsat fit, where a relatively large light quark mass $\sim 0.14\gev$ is preferred. This can be interpreted as an effective confinement requirement. The effect of a nonzero quark mass is to suppress dipoles larger than $\sim 1/m_\text{light}$, that in IPnonsat model have an unphysically large (unitarity violating) cross section. The final fit quality in both IPsat and IPnonsat models is similar, suggesting that, when describing the inclusive DIS data, the effective confinement effect in the IPnonsat has a comparable effect as the gluon saturation in the IPsat parametrization.

\begin{figure}[tb]
\begin{center}
\includegraphics[width=0.49\textwidth]{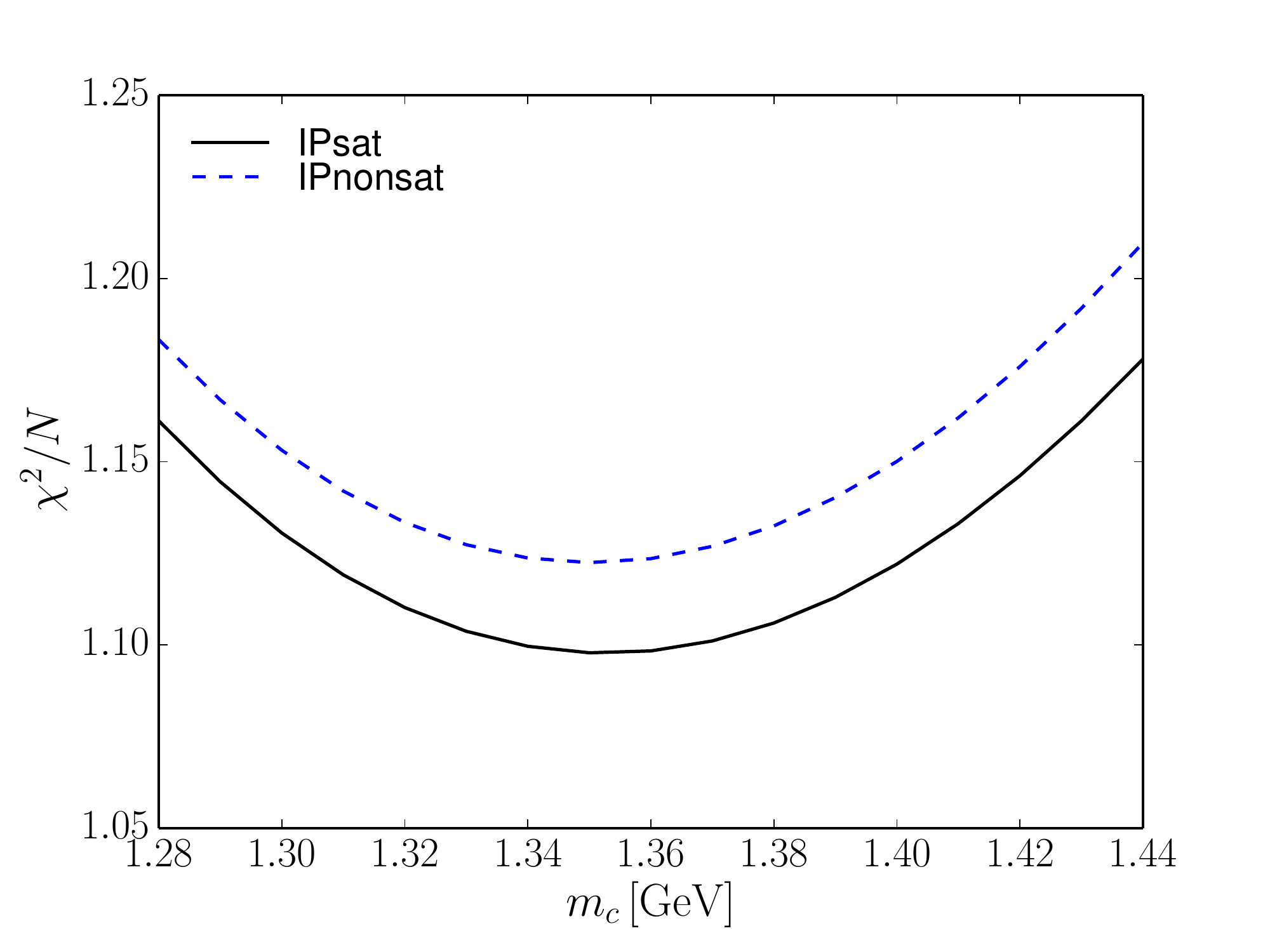}
\end{center}
\caption{
Fit quality to HERA inclusive~\cite{Aaron:2009aa} and charm~\cite{Abramowicz:1900rp} reduced cross section data as a function of the charm quark mass. In the IPsat fit the light quark mass is fixed to $0.03\gev$ (see text).
}\label{fig:charm_mass}
\end{figure}

Unlike previous works~\cite{Kowalski:2006hc,Rezaeian:2012ji} we find that the fit clearly prefers a charm quark mass $\sim 1.35\gev$, with the $\chi^2$ presenting a clear minimum. One main difference of our work with that of Ref.~\cite{Rezaeian:2012ji} is that the charm data are included in the fit, which allows us to constrain the charm mass simultaneously with the other parameters. The quality of the fit as a function of the charm mass is shown in Fig.~\ref{fig:charm_mass}, where all parameters are allowed to vary (except the light quark mass which is fixed to $0.03\gev$ in the IPsat model), while keeping $m_{c}$ constant.

\begin{figure*}[tb]
\begin{center}
\includegraphics[width=0.7\textwidth]{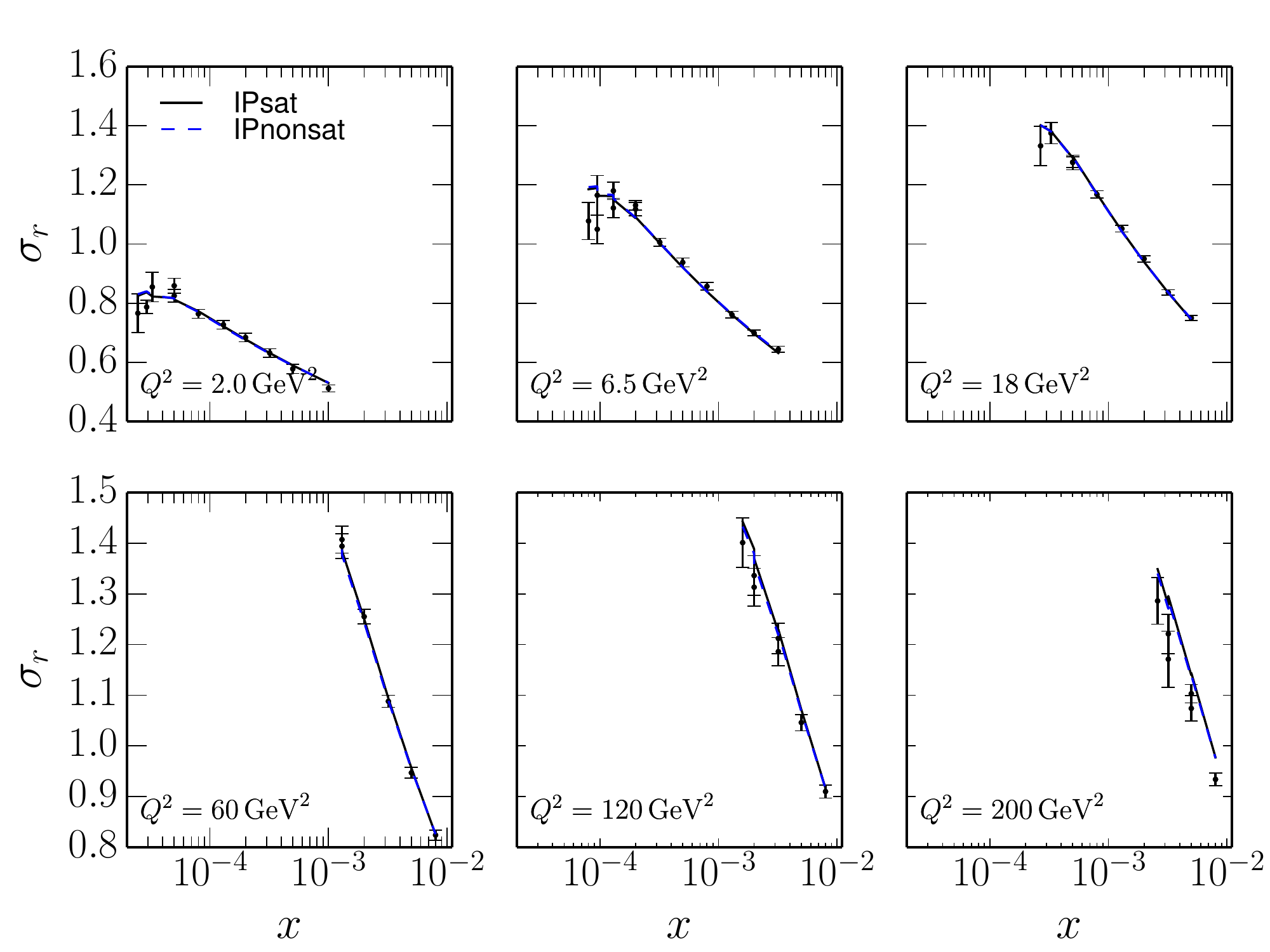}
\end{center}
\caption{
Inclusive reduced cross section from the fit including data up to $50\gev^2$, compared with the HERA data~\cite{Aaron:2009aa}.  }
\label{fig:total_sigmar}
\end{figure*}

\begin{figure*}[tb]
\begin{center}
\includegraphics[width=0.7 \textwidth]{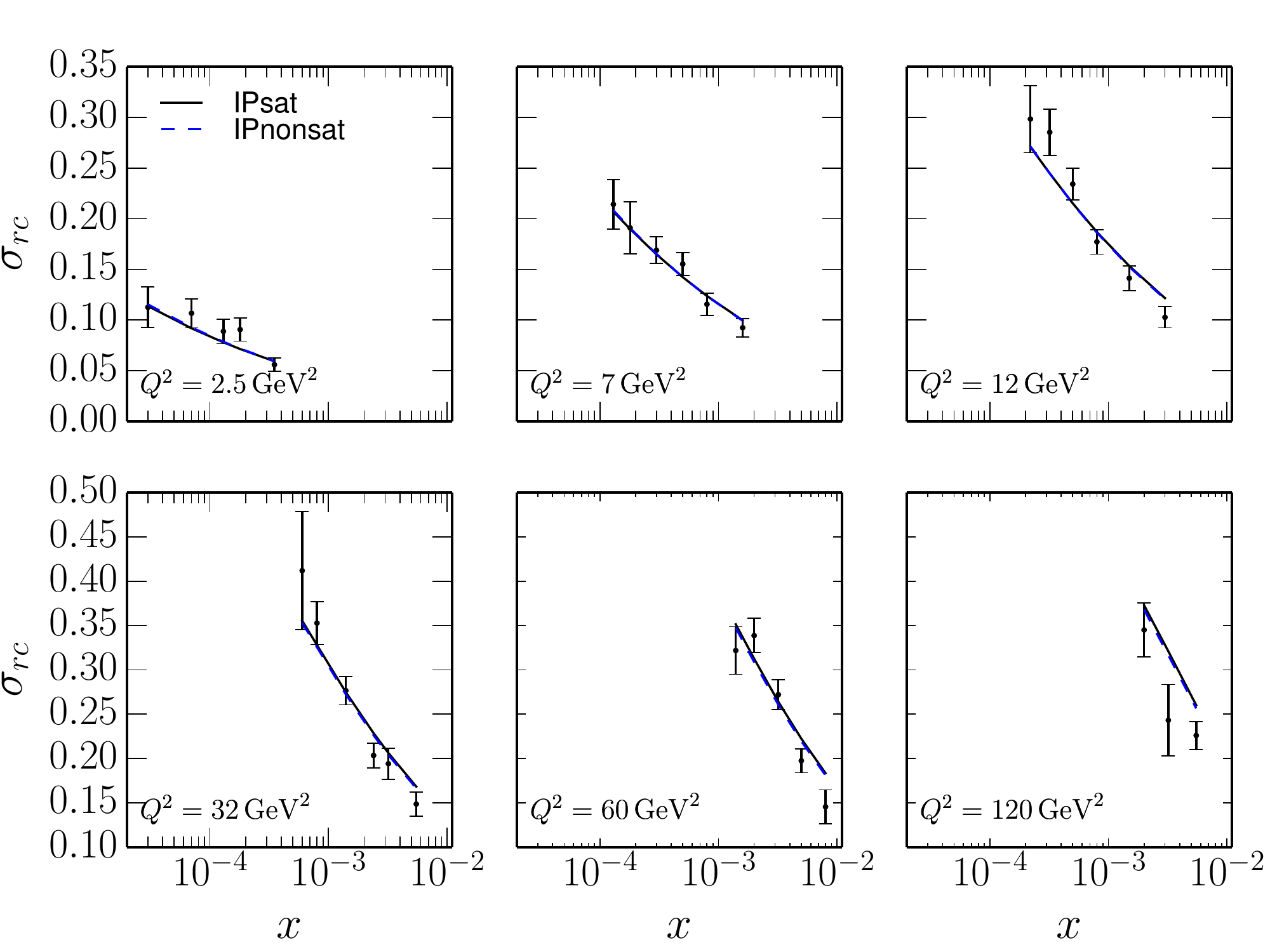}
\end{center}
\caption{
Charm reduced cross section calculated from the fit that includes data up to $50\gev^2$, compared with the HERA data~\cite{Abramowicz:1900rp}. }
\label{fig:charm_sigmar}
\end{figure*}

In Table \ref{tab:fits} we present the fitted parameters for two different values of $Q^{2}_\text{max}$. As can be seen from the $\chi^2$, the description of the precise HERA data is excellent, as already noted in previous works~\cite{Kowalski:2003hm,Kowalski:2006hc,Rezaeian:2012ji}. What is new here compared to the previous literature is the IPnonsat model parametrization, that we find to describe the combined HERA data equally well\footnote{In Ref.~\cite{Kowalski:2003hm} the IPnonsat model was fitted to older H1 and ZEUS data that have much larger uncertainties than the combined dataset used in this work}. This is demonstrated in Figs. \ref{fig:total_sigmar} and \ref{fig:charm_sigmar} where the reduced cross section and the charm contribution to it are shown and compared to the IPsat and IPnonsat model results, the curves being practically one on top of the other. We are also able to determine the quark masses from the fit. For the charm contribution, we note that there is some tension, the HERA data suggesting slightly slower $Q^2$ evolution that what is the outcome of our combined fit. This could be due to shortcomings of the model in describing the heavy quarks, as it happens in the collinear factorized framework where higher order corrections are needed for a proper description of the charm and bottom data~\cite{Ball:2017nwa}. It also might be influenced by the fact that the HERA collider was not particularly tuned to measuring heavy quarks, an issue that will be addressed in future colliders such as the EIC.

The fits done to HERA I and HERA I+II combined datasets result in comparable parameters (the largest difference being the scale parameter $C$, on which the results depend only logarithmically). Also, the newer dataset prefers a slightly smaller charm quark mass, but the differences are small. Thus the difference at the level of an observable will be negligible between the two different fits performed to different datasets. We will demonstrate this in Appendix~\ref{appendix:hera_i_ii}. We consider the top row of Table.~\ref{tab:fits} to be our main fit, as it relies on published datasets and only includes measurements in the kinematical domain where the applied model can be considered to be most reliable. The bottom quark reduced cross section included in the latest combination of HERA heavy quark data~\cite{H1:2018flt} is discussed in Appendix~\ref{appendix:bquark}.

\subsection*{Contribution from large dipoles}
As discussed above, especially in the IPnonsat model, a nonzero effective light quark mass is needed in order to obtain a satisfactory description of the HERA data. This means that the reduced cross section data is sensitive to the contributions from (possibly non-perturbatively) large dipoles, whose formation should be suppressed by confinement scale effects. In the IPsat parametrization, the imposed unitarity requirement limits the scattering probability not to exceed unity at large dipoles, but large dipoles can still have a numerically significant contribution.

\begin{figure}[tb]
\begin{center}
\includegraphics[width=0.49\textwidth]{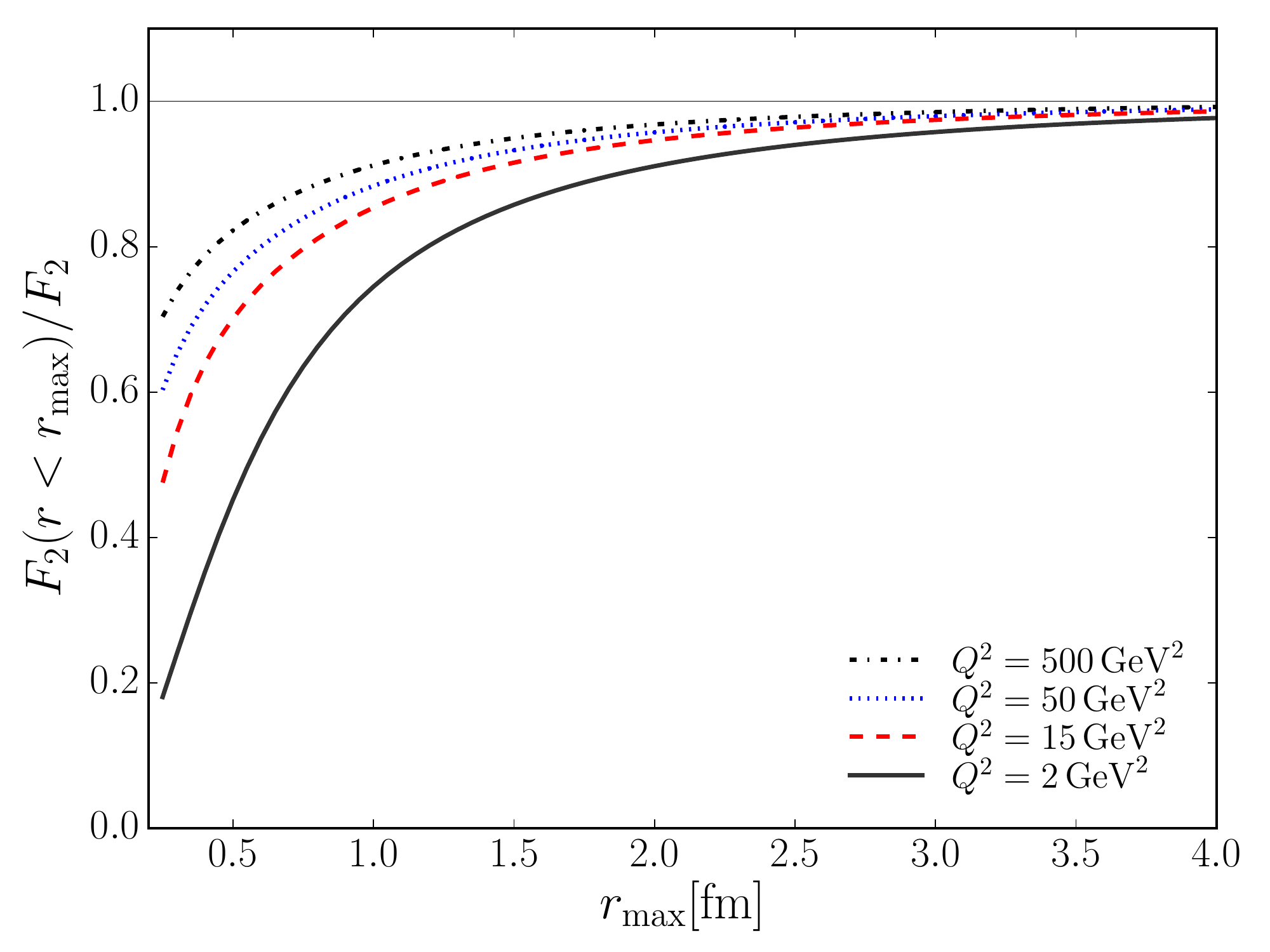}
\end{center}
\caption{
Contribution to $F_2$ at $x=0.01$ from the IPsat model from dipoles smaller than $r_\text{max}$ at different $Q^2$. Results for IPnonsat are similar. }
\label{fig:f2_maxr}
\end{figure}

\begin{figure}[tb]
\begin{center}
\includegraphics[width=0.49\textwidth]{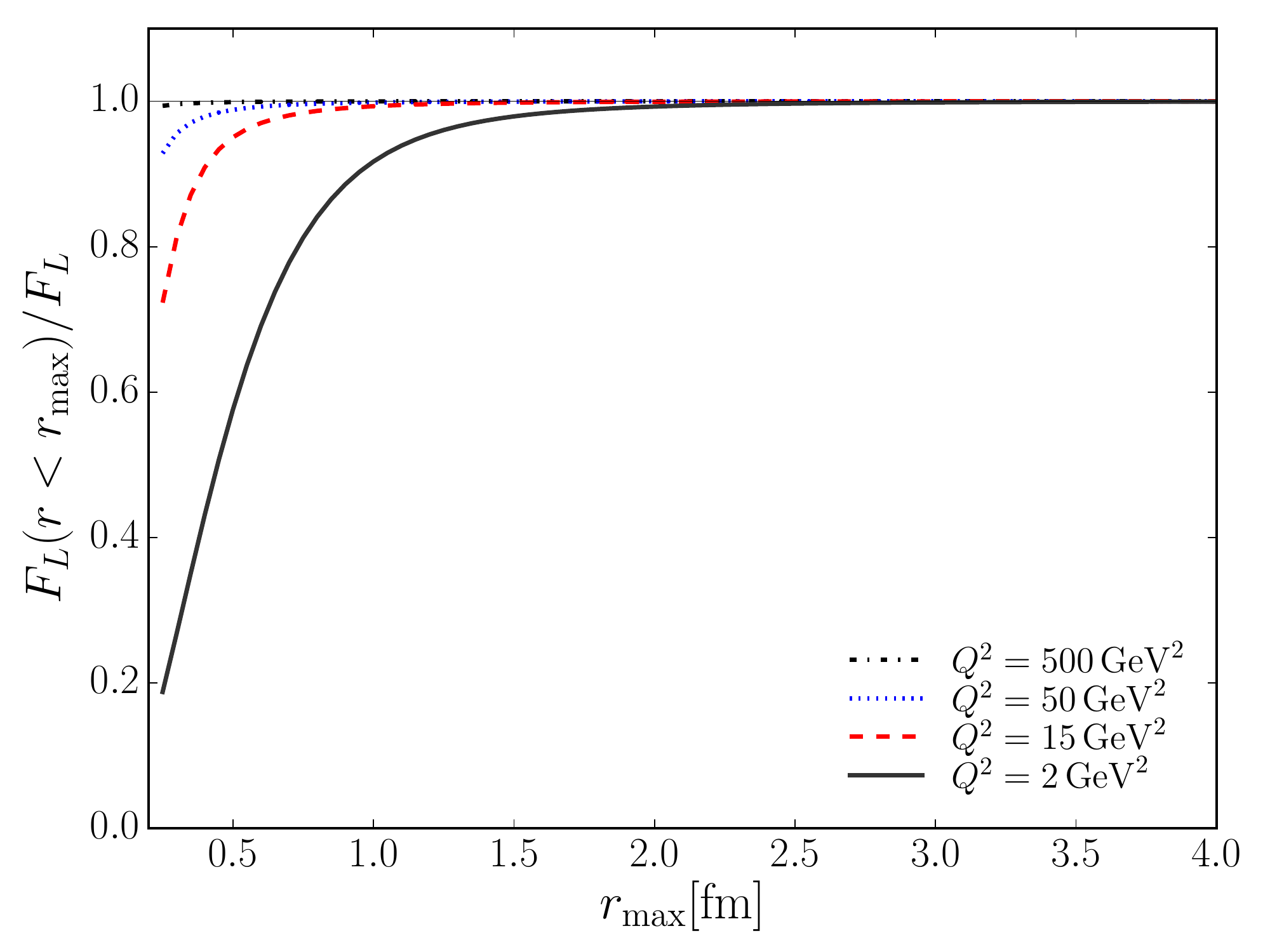}
\end{center}
\caption{
Contribution to $F_L$ at $x=0.01$ from dipoles smaller than $r_\text{max}$ at different $Q^2$.  }
\label{fig:fl_maxr}
\end{figure}

The fractional contribution from large dipoles to the total $F_2$ and $F_L$ is shown in Figs.~\ref{fig:f2_maxr} and \ref{fig:fl_maxr}, respectively. These calculations are done using the IPsat fit (first line of Tab.~\ref{tab:fits}), using the IPnonsat fit would result in very similar $r_\text{max}$ dependence. For $F_2$, even at relatively large $Q^2 \sim 500\gev^2$, 10\% of the total structure function originates from dipoles larger than $1 \fm$. On the other hand, the HERA reduced cross section data have relative uncertainties at the percentage level, much smaller than the contribution we obtain from the non-perturbatively large dipoles.

The reason for this large contribution is that there is a large so called \emph{aligned jet} contribution: in the limits $z\to 0$ and $z\to 1$ the large dipole contribution to the transverse photon-proton cross section is only suppressed by the light quark mass as $\sim e^{-m_\text{light} r}$. This can be seen from the virtual photon wave function, Eq.~\eqref{eq:photon_t}. On the other hand, as can be seen from Fig.~\ref{fig:fl_maxr}, in the case of $F_L$ at moderate $Q^2$ the contribution from the region $r\gtrsim 1\fm$ is negligible. This is due to the extra factor $z^2(1-z)^2$ in the longitudinal photon wave function, Eq.~\eqref{eq:photon_l}, which suppresses the endpoint contributions. Thus, we would prefer to fit the $F_L$ data instead of the reduced cross section measurements which is dominated by $F_2$. However, the HERA $F_L$ data~\cite{Aaron:2008ad,Chekanov:2009na} are not precise enough for a detailed comparison with the dipole model calculations. 

\begin{figure}[tb]
\begin{center}
\includegraphics[width=0.49\textwidth]{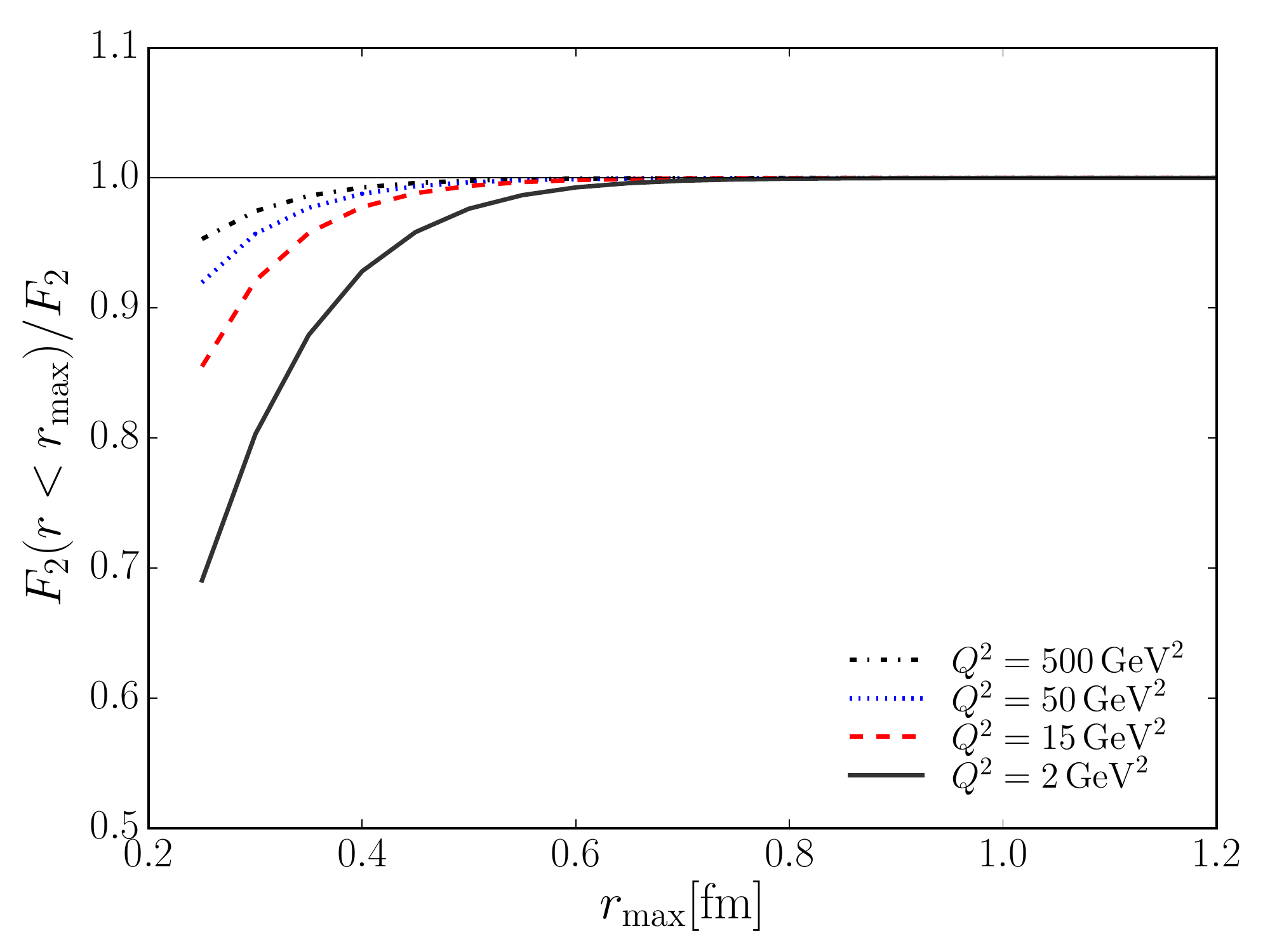}
\end{center}
\caption{
Contribution to charm structure function $F_{2,c}$ at $x=0.01$ from dipoles smaller than $r_\text{max}$ at different $Q^2$. Note that the scales are different than in Figs.~\ref{fig:f2_maxr} and \ref{fig:fl_maxr}.}
\label{fig:f2_maxr_charm}
\end{figure}

Future DIS facilities EIC and LHeC have plans to measure proton structure functions (including $F_L$) at an unprecedented accuracy~\cite{AbelleiraFernandez:2012cc,Accardi:2012qut,Aschenauer:2017jsk}. Similarly, studying only the charm contribution to the total cross section limits the contribution from large dipoles as demonstrated in Fig.~\ref{fig:f2_maxr_charm}. In case of the $F_{2,\text{charm}}$, even at small $Q^2$ contribution from dipoles larger than $\sim 0.6\fm$ is negligible (but very small dipoles are not sensitive to the saturation effects either). In general we find that $F_2$, $F_L$ and $F_{2,\text{charm}}$ are sensitive to different dipole sizes, and future DIS data covering all these structure functions will provide much more precise constraints.

\begin{figure}[tb]
\begin{center}
\includegraphics[width=0.49\textwidth]{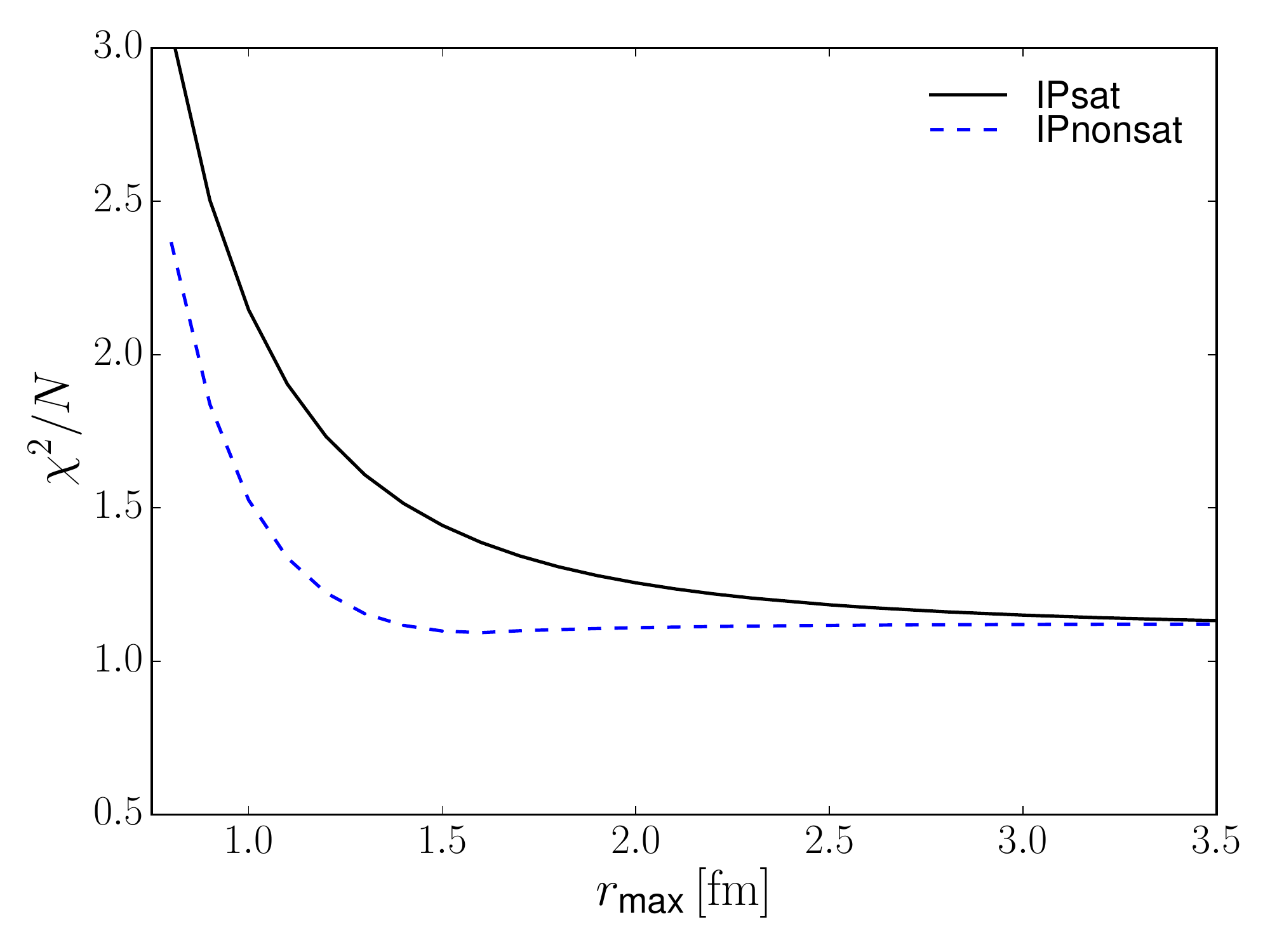}
\end{center}
\caption{
Fit quality with different cutoffs for large dipoles. In the IPsat model the light quark mass is fixed to $m_\text{light}=0.03\gev$.  }
\label{fig:chisqr_maxr}
\end{figure}

\begin{figure}[tb]
\begin{center}
\includegraphics[width=0.49\textwidth]{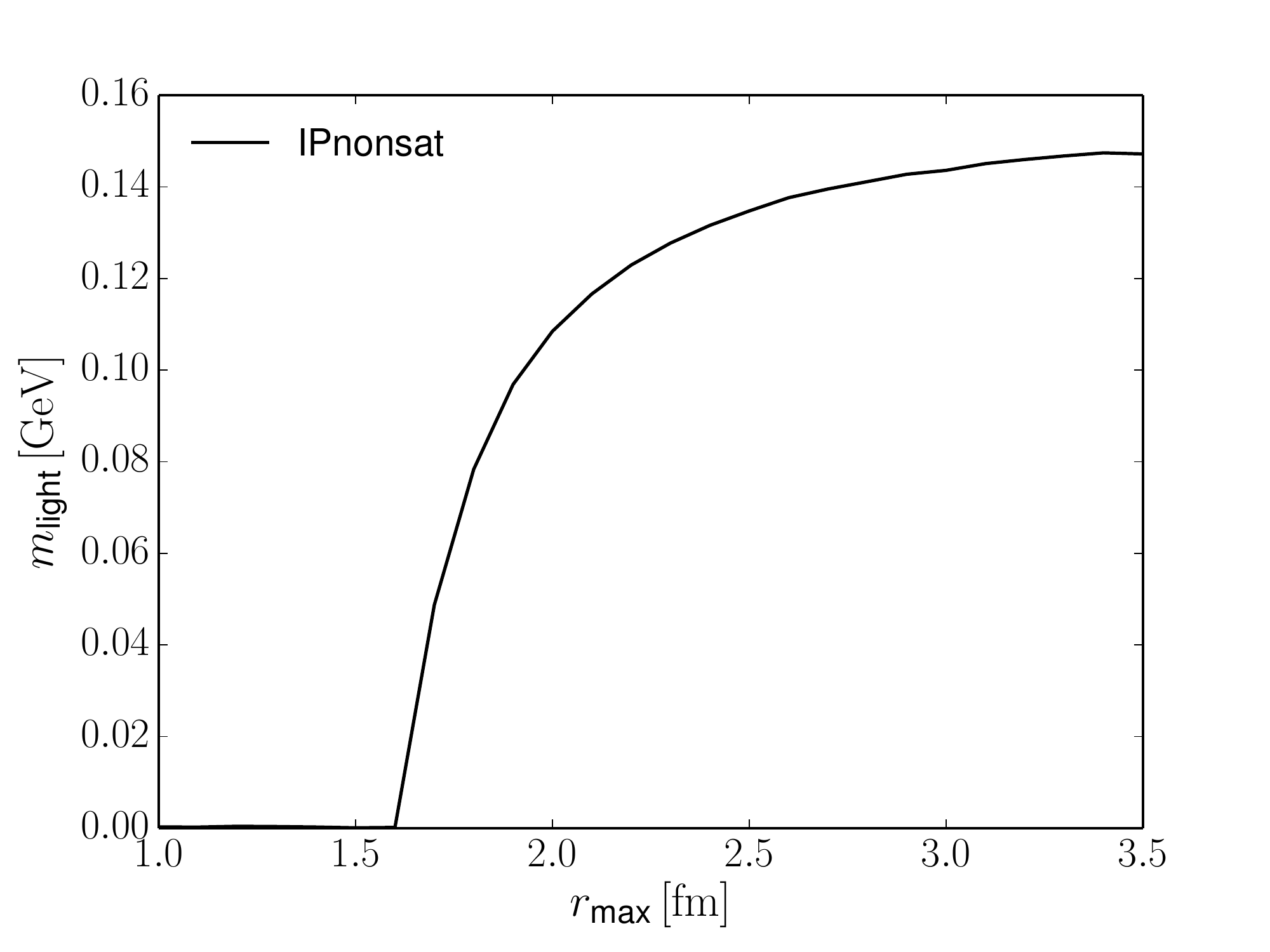}
\end{center}
\caption{
Light quark mass obtained as a fit result with the IPnonsat model as a function of the infrared cut for the large dipoles.  }
\label{fig:rmax_lightmass}
\end{figure}

In order to further study how much large dipoles affect the fit result, we perform the fits to HERA I inclusive and charm cross section data up to $Q_\text{max}^2=50\gev^2$ with different cutoffs for large dipoles $r_\text{max}$. The resulting fit quality is shown in Fig.~\ref{fig:chisqr_maxr}. We find that in order to obtain a good fit to the HERA data, we have to include dipoles up to $\sim 2 \dots 2.5 \fm$ in the IPsat model. In the case of the IPnonsat parametrization, the fit can compensate the effect of the $r_\text{max}$ cutoff as the light quark mass is also a fit parameter, thus the fit quality is more stable with respect to the infrared cutoff. The IPnonsat fit drives the light quark mass to zero when $r_\text{max}$ becomes $\sim 1.6\fm$, and it is not possible to fit the HERA data with a much smaller cutoff. The dependence of the light quark mass on $r_\text{max}$ is shown in Fig.~\ref{fig:rmax_lightmass} which further demonstrates that in the IPnonsat model the inclusion of large dipoles requires a larger light quark mass to suppress the contribution from this unphysical region. In the IPsat model, the fits prefer a zero light quark mass at all $r_\text{max}$.

\section{Dipole scattering amplitude}
\label{amplitude}
\begin{figure}[tb]
\begin{center}
\includegraphics[width=0.49\textwidth]{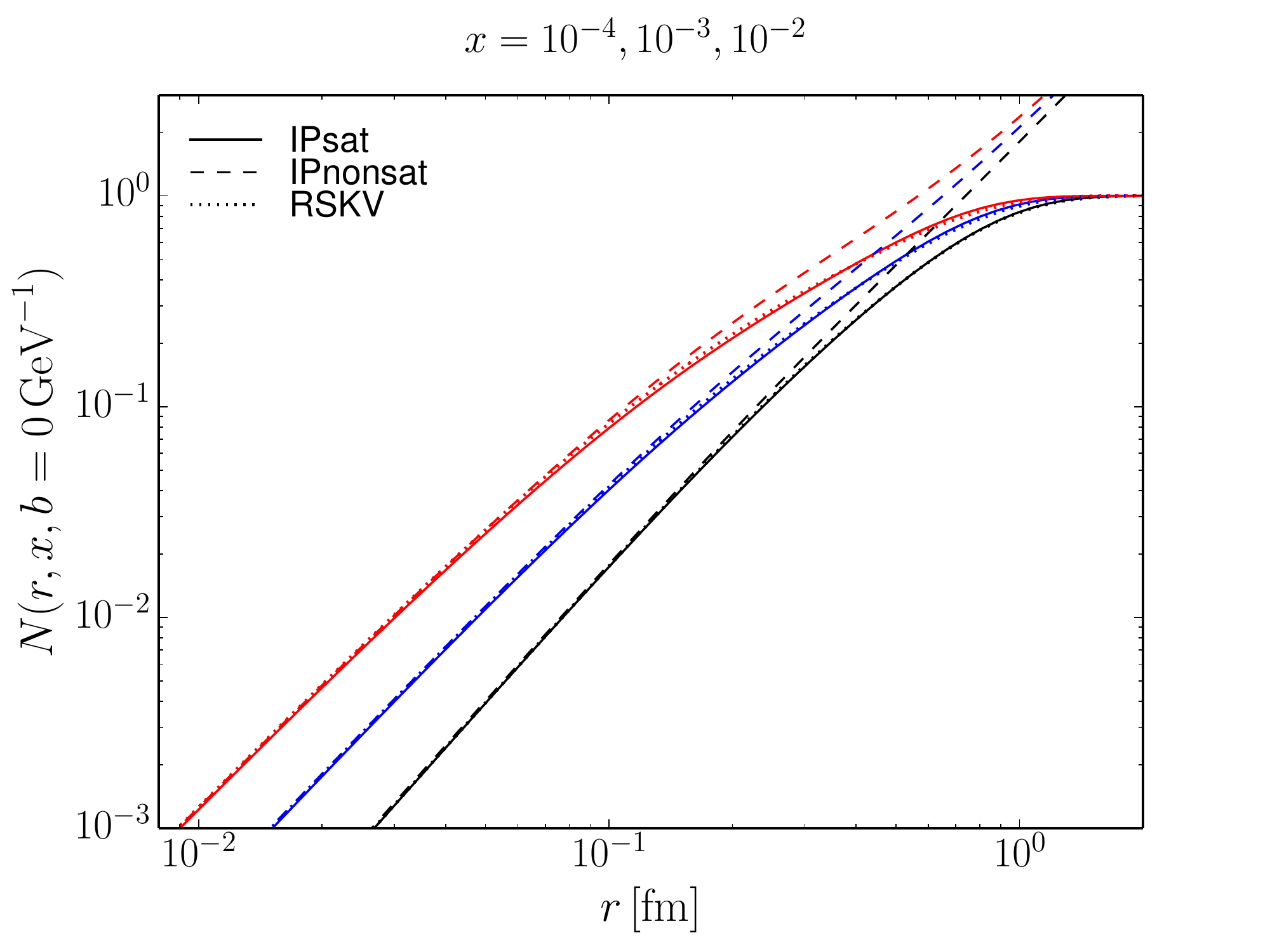}
\end{center}
\caption{
The obtained dipole amplitudes at $x=10^{-6}$ (red), $x=10^{-4}$ (blue) and $x=10^{-2}$ (black). RSKV refers to the previous fit~\cite{Rezaeian:2012ji}.
}\label{fig:dipoleamplitude}
\end{figure}

The resulting dipole amplitude at $b=0$ is shown in Fig.~\ref{fig:dipoleamplitude}, where we compare it with the previous results from \cite{Rezaeian:2012ji} labeled as ``RSKV''. Even thought our study has incorporated some refinements (variable flavor number scheme in the DGLAP evolution, quark masses determined by the fit, inclusion of the charm reduced cross section data), the base model is essentially the same and, by fitting similar data sets one is expected to obtain compatible dipole amplitudes, despite some numerical differences in the fitted parameters. In the IPnonsat model the evolution at the initial scale in $x$ is very slow ($\lambda_g$ being close to $0$), thus there is basically no evolution at large $r$, in the region where the IPsat parametrization has already reached unity (and where the IPnonsat model gives unphysical results). 

\begin{figure}[tb]
\begin{center}
\includegraphics[width=0.49\textwidth]{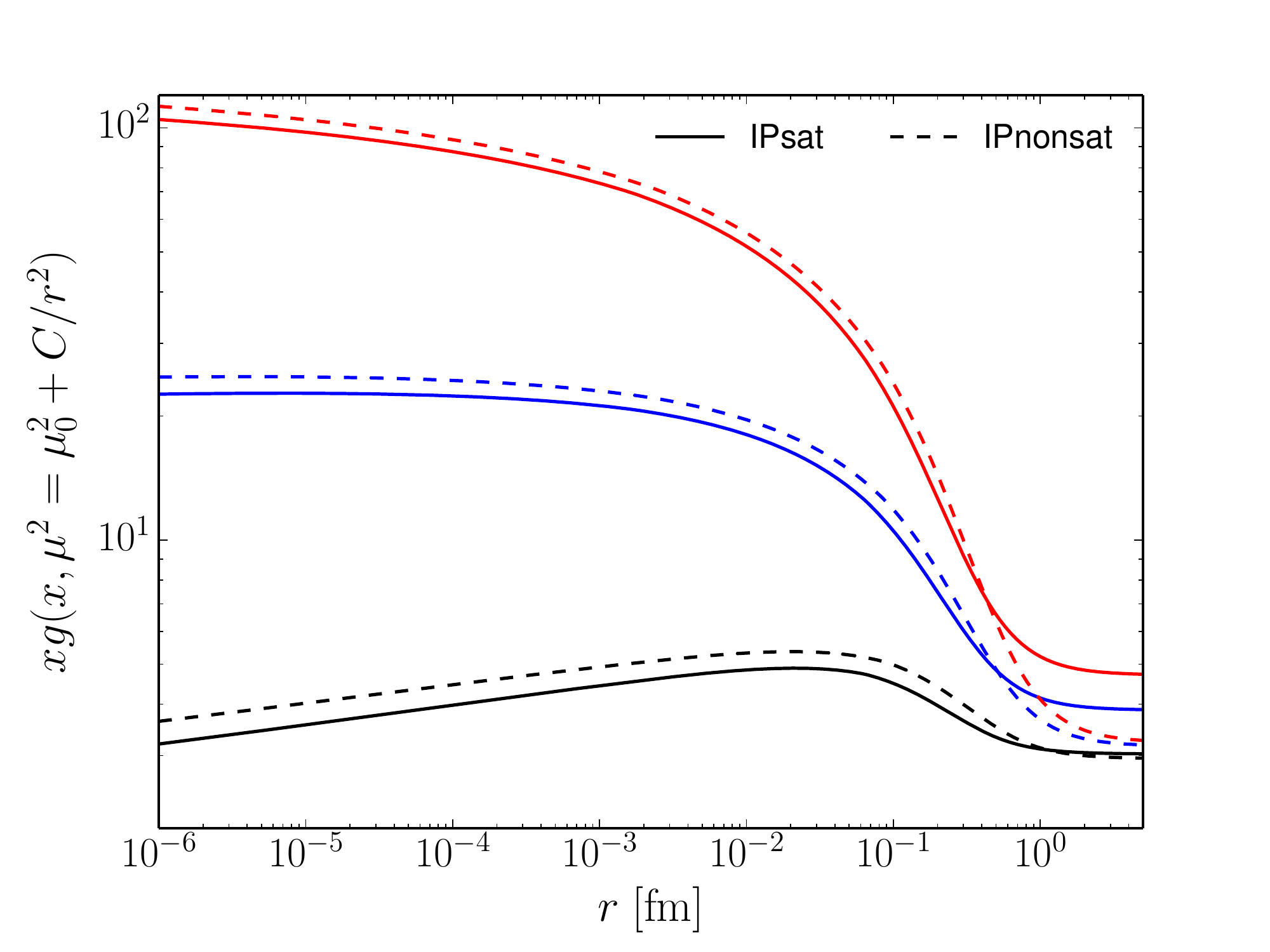}
\end{center}
\caption{
Gluon density $xg$ as a function of the dipole size $r$ for $x=10^{-4}$ (red), $x=10^{-3}$ (blue) and $x=10^{-2}$ (black) from top to bottom. 
 }\label{fig:xg}
\end{figure}

To demonstrate the evolution of the gluon distribution function we plot $xg(x,\mu^2=\mu_0^2+C/r^2)$ in Fig.~\ref{fig:xg} as a function of $r$ using both the IPsat and IPnonsat fitted parameters to initialize the DGLAP evolution. At large scales the two parametrizations have small differences, as the effect of the initial condition is washed out in the evolution. Close to the initial scale $\mu_0^2$ there is basically no evolution if the IPnonsat model parametrization is used ($\lambda_g \approx 0$, which forces the dipole scattering amplitude not to grow in the region where it is already violating unitarity), unlike in the case of the IPsat model.

At large scales and at sufficiently large $x \gtrsim 10^{-3}$ it is also visible that the gluon density starts to decrease as the scale $\mu^2 \sim C/r^2$ increases. This is expected, as the momentum conservation in the DGLAP evolution removes the larger-$x$ gluons as they are splitting into the smaller-$x$ ones. Similar results were already found in Ref.~\cite{Kowalski:2003hm}. This effect is strong close to $x\sim 10^{-2}$, where the decreasing gluon density is probed already by dipoles that have large enough sizes to contribute significantly on $F_2$.

\begin{figure}[tb]
\begin{center}
\includegraphics[width=0.49\textwidth]{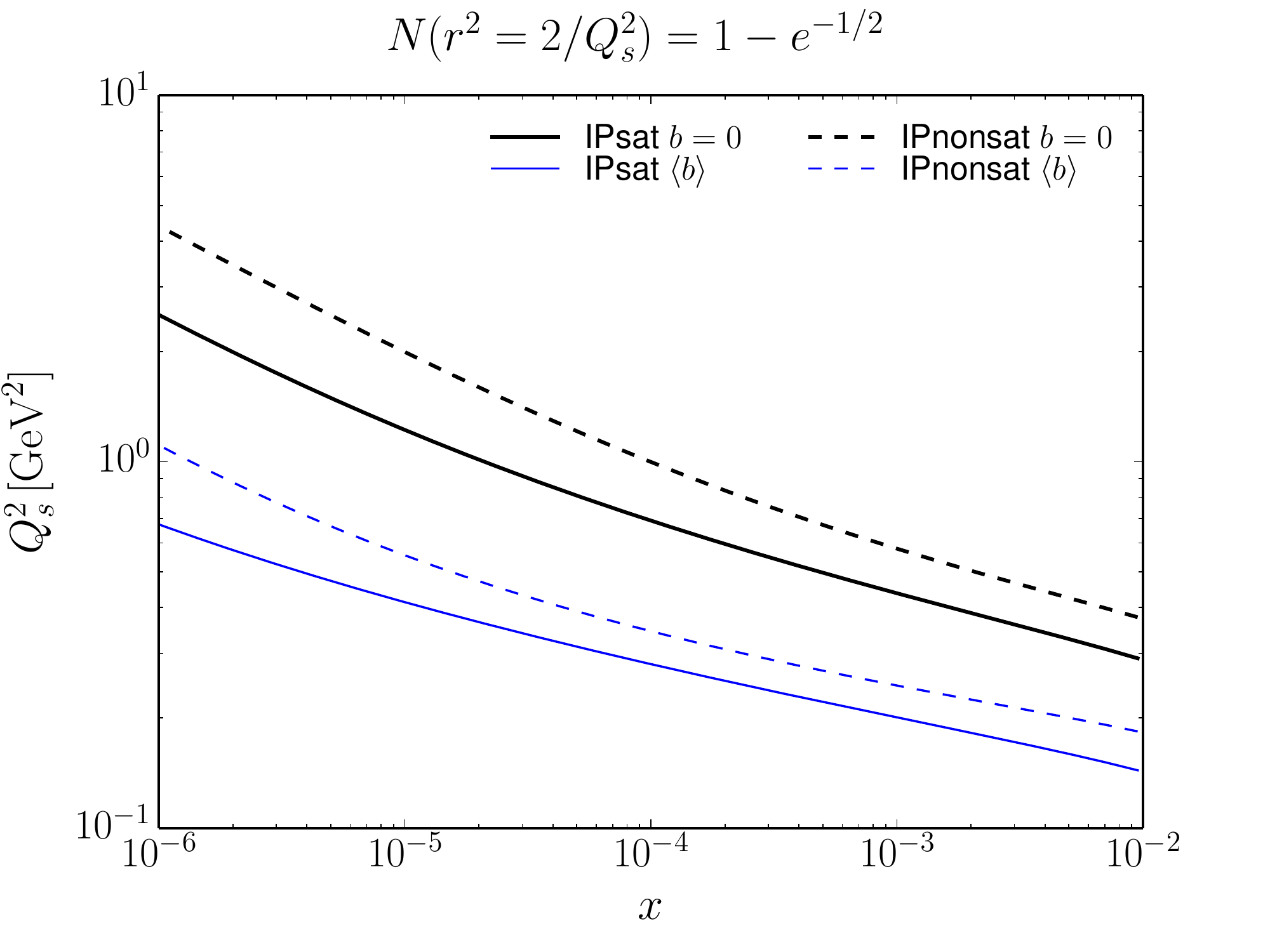}
\end{center}
\caption{
Saturation scale at the center (thick black lines) of the proton and at average impact parameter $\langle b \rangle \approx 0.46\fm$ (thin blue lines).
}\label{fig:qs}
\end{figure}

The point at which the non-linear effects become relevant is characterized by the saturation scale $Q_s^2$. To determine it we use the definition
\begin{equation}
\label{eq:satscale}
N(r^2=2/Q_s^2,x,b) = 1-e^{-1/2}.
\end{equation}
The extracted saturation scale as a function of $x$ is shown in Fig.~\ref{fig:qs}. Here, $Q_s^2$ is extracted at the central impact parameter $b=0$, and at the average $\langle b \rangle$ defined such that 
\begin{equation}
\int_0^{\langle b \rangle}  \der b \, b \, T_p(b) = \frac{1}{2} \int_0^\infty  \der b \, b \, T_p(b).
\end{equation}
This definition gives $\langle b \rangle \approx 0.46\fm$. The difference between the IPsat and IPnonsat parametrizations remains small at all values of Bjorken-$x$, the IPnonsat model having in general slightly faster evolution. As expected based on previous analyses (e.g.~\cite{Rezaeian:2012ji,Lappi:2013zma}), the saturation scale of the proton is at the $\lqcd$ range in the region $x\sim 10^{-2}$, and the region of $Q_s^2$ being perturbative is reached below $x \lesssim 10^{-4} \dots 10^{-5}$ (note that the absolute value of $Q_s^2$ depends on the definition \eqref{eq:satscale}).

\section{Exclusive vector meson production}
\label{EVM}
Additional information about the proton structure can be obtained by studying exclusive processes. They are particularly powerful in probing the gluonic structure, as at leading order in collinear factorization the vector meson production cross section is proportional to the \emph{squared} gluon distribution~\cite{Ryskin:1992ui}. In addition to being sensitive to the total gluonic density, in exclusive process the momentum transfer $\Delta = \sqrt{-t}$ is the Fourier conjugate to the impact parameter, which makes it possible to probe also the impact parameter dependence.

In the dipole picture, the scattering amplitude for exclusive vector meson production reads (see e.g. Ref.~\cite{Kowalski:2006hc})
\begin{multline}
\A^{\gamma^* p \to V p} = \int \der^2 \rt \der^2 \bt \frac{\der z}{4\pi} (\Psi^* \Psi_V)(\rt, Q^2,z)  \\
\times e^{-i \bt \cdot {\bf \Delta}} \frac {\ud \sigmadip}{\ud^2 \bt}.
\end{multline}
This expression has a straightforward interpretation. First, the incoming virtual photon splits into a quark-antiquark pair as described by the virtual photon wave function $\Psi$. The dipole then scatters elastically off the proton with cross section $\sigma_\text{dip}$, and ultimately forms the final state vector meson described by the wave function $\Psi_V$. The scattering amplitude in the momentum space is obtained by calculating the Fourier transform from the coordinate space, with the momentum transfer ${\bf \Delta}$ being the Fourier conjugate to the impact parameter $\bt$. 
Here we have neglected the off-forward correction to the vector meson wave function~\cite{Bartels:2003yj}.

The exclusive vector meson production cross section reads
\begin{equation}
\frac{\der \sigma^{\gamma^* p \to Vp}}{\der t} = \frac{1}{16\pi} \left| \A^{\gamma^* p \to V p}  \right|^2.
\end{equation} 
In addition, we include the corrections due to the real part of the scattering amplitude neglected when deriving the above result, and the so called ``skewedness correction" which takes into account the fact that in the two-gluon exchange the two gluons carry different amount of longitudinal momentum~\cite{Shuvaev:1999ce}. These corrections are included as in Ref.~\cite{Lappi:2010dd} and, to a good approximation, only affect the overall normalization of the diffractive cross section. 

Unlike the virtual photon wave function used to calculate inclusive cross sections, the vector meson wave function can not be calculated perturbatively. We use here the Boosted Gaussian parametrization as in Ref.~\cite{Kowalski:2006hc}, where one assumes that the vector meson is a quark-antiquark state with spin and polarization structure the same as in the case of the photon. This assumption makes it possible to write the overlap between the vector meson $V$ and the virtual photon wave function in case of the transverse polarization as 
\begin{multline}
(\Psi^*_V\Psi)_T = \hat e_f e \frac{\nc}{\pi z(1-z)} \left\{ m_f^2 K_0(\varepsilon r) \phi_T(r,z) \right. \\
\left. - [z^2+(1-z)^2] \varepsilon K_1(\varepsilon r) \partial_r \phi_T(r,z) \right\},
\end{multline}
and for the longitudinal polarization
\begin{multline}
(\Psi^*_V\Psi)_L =\hat e_f e \frac{\nc}{\pi} 2 Q z(1-z) K_0(\varepsilon r) \\
	\times \left[ M_V \phi_L(r,z) +  \frac{m_f^2 - \nabla^2}{M_V z(1-z)} \phi_L(r,z) \right].
\end{multline}
The scalar part of the vector meson is parametrized as
\begin{multline}
\phi_{T,L}(r,z) = N_{T,L} z(1-z) \exp \left( -\frac{m_f^2 R^2}{8z(1-z)} \right. \\
 - \frac{2z(1-z)r^2}{R^2} \left. + \frac{m_f^2 R^2}{2} \right).
\end{multline}
The advantage of this parametrization is that the wave function has the proper short-distance behavior $\sim z(1-z)$ in the limit of massless quarks. The normalization factors $N_{T,L}$ and the width $R$ are fixed by requiring that the decay width to the electron channel (calculated using the longitudinal polarization as in Ref.~\cite{Kowalski:2006hc}) reproduces the experimental value, and that the wave function is properly normalized. As these parameters depend on the quark masses, we calculate them for $J/\Psi$, $\rho$ and $\phi$ for the same values obtained in the fits to inclusive data for the IPsat and IPnonsat parametrizations. The obtained values and the comparison with the results from \cite{Kowalski:2006hc} are shown in Table.~\ref{table:boostedgaussian}. We remark that if one were to calculate the decay width using the transverse polarization, the final numbers would be slightly different as noted in Ref.~\cite{Kowalski:2006hc}.

\begin{table*}[tb]
  \centering
  \begin{tabular}{ccccccc}
    \toprule
    Meson  &  $M_V$ [GeV]  & Type & $m_f$ [GeV] & $R$ $[\mathrm{GeV}^{-1}]$ & $N_T$ & $N_L$   \\
    \hline
    $J/\Psi$ & 3.097 & IPsat & 1.3528 & 1.5070 & 0.5890 & 0.5860  \\
     $J/\Psi$ & 3.097 &  IPnonsat & 1.3504 & 1.5071 & 0.5899  & 0.5868 \\
    $J/\Psi$ & 3.097 & KMW  & 1.4 & 1.5166 & 0.578 & 0.575 \\
     \hline
     $\phi$ & 1.019 & IPsat & 0.03 & 3.3922 & 0.9950 &  0.8400 \\
     $\phi$ & 1.019 & IPnonsat & 0.1516 &  3.3530 & 0.9072  & 0.8196  \\
     $\phi$ & 1.019 & KMW  & 0.14 & 3.347 & 0.919  & 0.825 \\
    \hline
     $\rho$ &  0.776 &IPsat & 0.03 & 3.6376 &  0.9942 & 0.8926  \\
    $\rho$ & 0.776 &  IPnonsat & 0.1516 & 3.5750 &  0.8978  & 0.8467 \\
     $\rho$ & 0.776 & KMW & 0.14 & 3.592 & 0.911  & 0.853   \\
     
 \end{tabular}

 \caption{Parameters for the Boosted Gaussian wave function corresponding to the quark masses obtained for the IPsat and IPnonsat parametrizations. For comparison we include the results from \cite{Kowalski:2006hc} (labeled as KMW), also determined using the longitudinal polarization. }
  \label{table:boostedgaussian}
 \end{table*}

 \begin{figure}[tb]
\begin{center}
\includegraphics[width=0.49\textwidth]{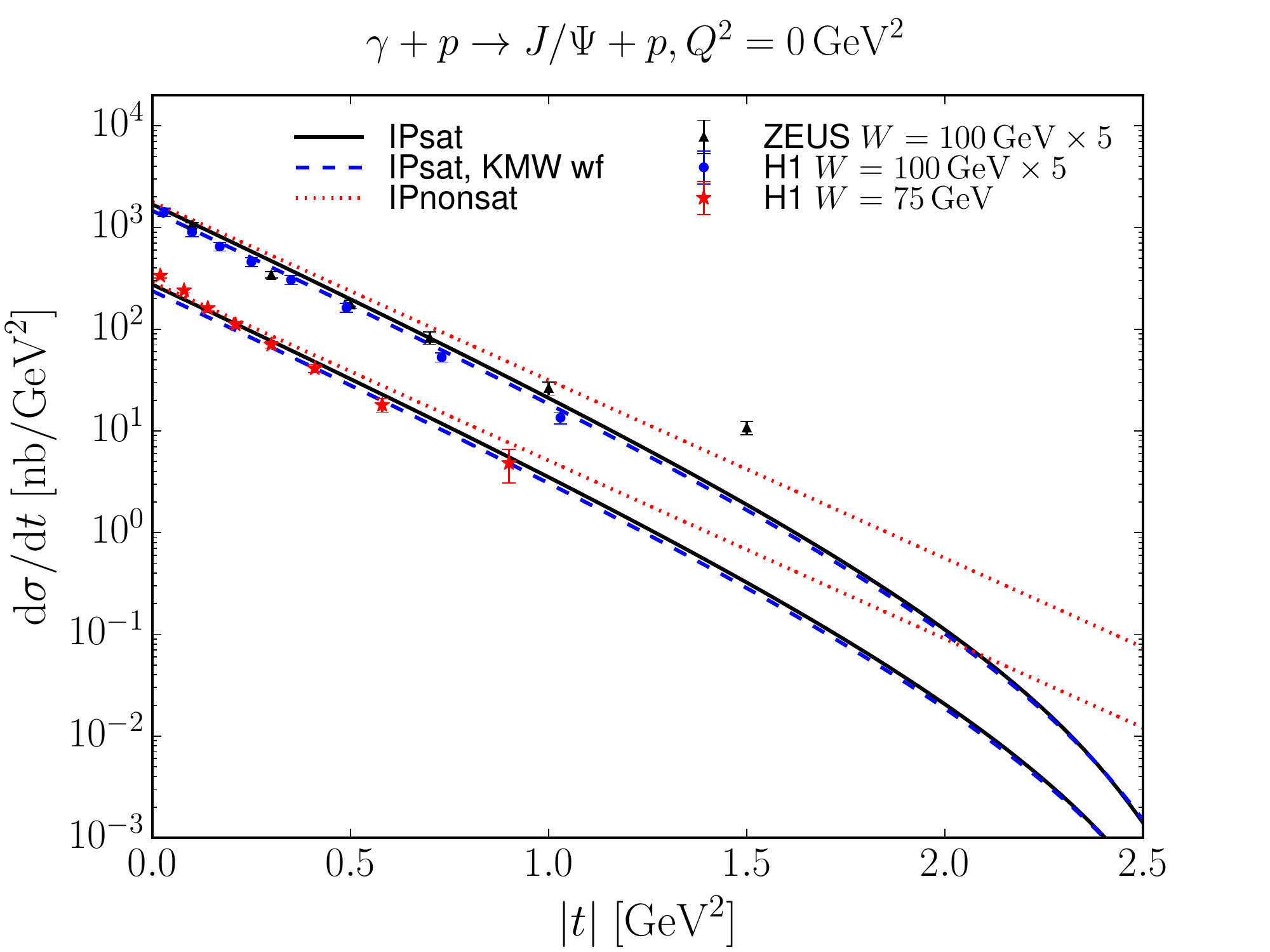}
\end{center}
\caption{
Differential J/$\Psi$ photoproduction cross section as a function of momentum transfer $-t={\bf \Delta}^2$ at two different center-of-mass energies. The dashed line is obtained by using the wave function provided in Ref.~\cite{Kowalski:2006hc} where the charm quark mass is $m_c=1.4\gev$. The other IPsat and IPnonsat curves use the wave function parametrizations from Table~\ref{table:boostedgaussian}. The $W=75\gev$ data are from Ref.~\cite{Alexa:2013xxa} and the $W=100\gev$ data from Refs.~\cite{Aktas:2005xu,Chekanov:2002xi}. The high-energy results are scaled by $5$ for illustrational purposes. 
}\label{fig:jpsi_spectra}
\end{figure}

 \begin{figure}[tb]
\begin{center}
\includegraphics[width=0.49\textwidth]{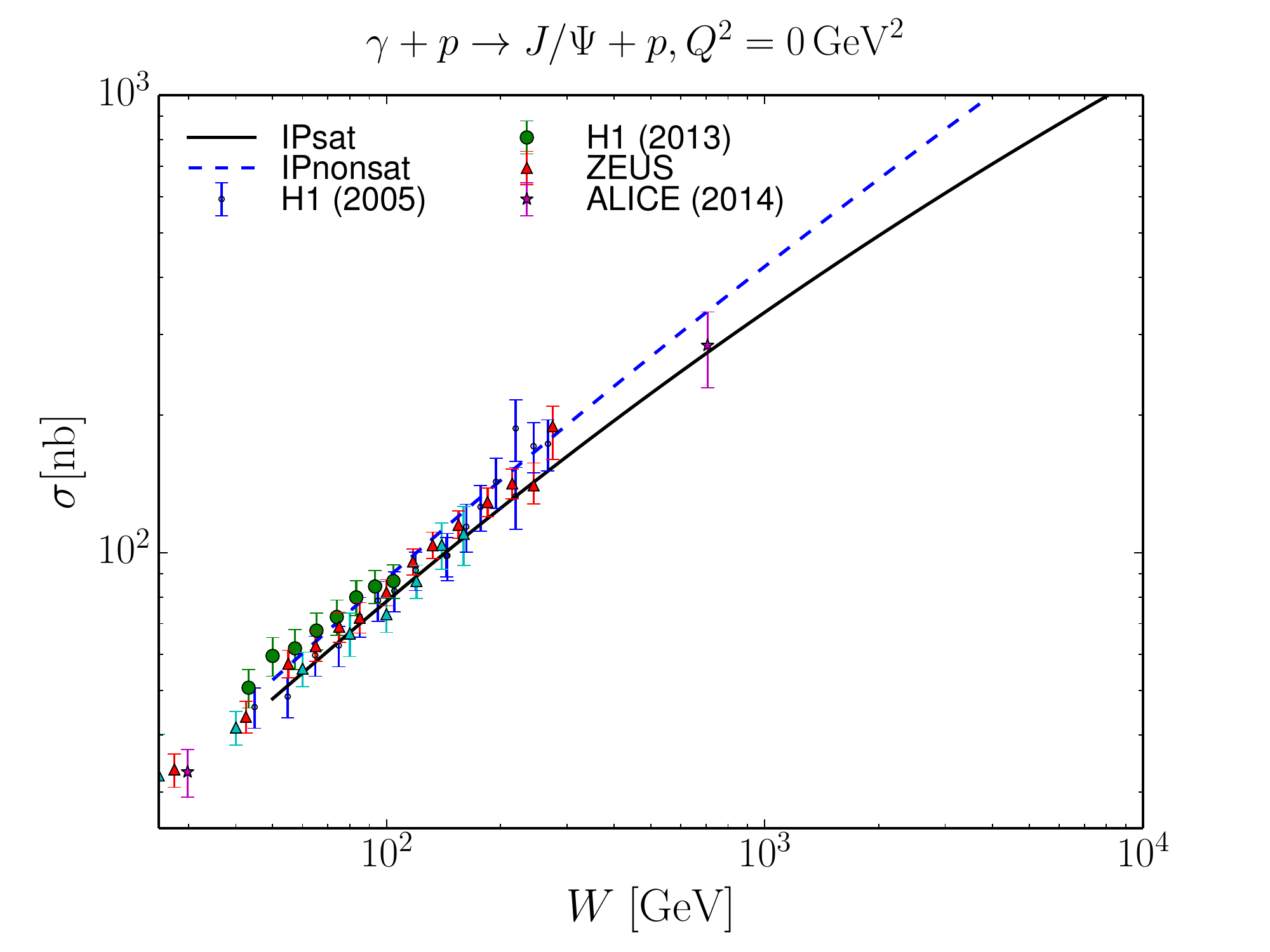}
\end{center}
\caption{
Total exclusive $\gamma p \to J/\Psi p$ production cross section as a function of $W$. 
}\label{fig:jpsi_totxs}
\end{figure}

Due to the small light quark mass especially in the IPsat model parametrization, the photoproduction of $\rho$ and $\phi$ mesons can not be reliably calculated from our model. In the case of J/$\Psi$, the charm quark mass provides the necessary large scale that cuts out large dipoles, and makes it possible to calculate exclusive J/$\Psi$ production down to $Q^2=0 \gev^2$. This is advantageous, as recently it has become possible to measure exclusive vector meson photoproduction in ultraperipheral collisions at RHIC and the LHC~\cite{Bertulani:2005ru}.

In the literature there are some inconsistencies using the vector meson wave function parametrization from Ref.~\cite{Kowalski:2006hc} together with dipole model fits with different choices for the charm quark mass. In order to quantify the effect of having a consistent quark mass in the dipole model fit and in the vector meson wave function, we show in Fig.~\ref{fig:jpsi_spectra} the J/$\Psi$ production cross section using our IPsat fit (where $m_c\approx1.35\gev$) and the widely used wave function from Ref.~\cite{Kowalski:2006hc} (where $m_c=1.4\gev$, referred as KMW). The larger quark mass in the KMW parametrization reduces the cross section by approximately 14\%. We note that the uncertainties related to the modeling of the vector meson wave function are larger than this, see e.g. Refs.~\cite{Kowalski:2006hc,Lappi:2013am}. The IPsat and IPnonsat results are practically on top of each other at small $|t|$.
The agreement with the HERA data is good, except that we can not reproduce the small change of the $t$ slope at $|t| \lesssim 0.1\gev^2$ visible in the $W=75\gev$ data\footnote{Which is described accurately in the IP-Glasma model calculation in Refs.~\cite{Mantysaari:2016ykx,Mantysaari:2016jaz}}.

At large $|t|$ the different form factors generate different spectra. The Fourier transform of the IPnonsat dipole amplitude is exactly Gaussian, and the spectra goes like $e^{-B_p |t|}$. In the IPsat parametrization, the proton density profile is actually $\sim \exp(-e^{-b^2/2B_p})$, thus its Fourier transform is a more complicated function which generates diffractive dips at large $-t$. At $W\sim 100\gev$, we get the location for the first diffractive minimum to be $|t|\sim 2.5\gev^2$ (see also Ref.~\cite{Armesto:2014sma} for discussion about the energy dependence of the dip location). 

The total $J/\Psi$ production cross section, calculated using our IPsat and IPnonsat model fits, is shown in Fig.~\ref{fig:jpsi_totxs}. The results are compared with the HERA data from the H1~\cite{Aktas:2005xu,Alexa:2013xxa} and ZEUS~\cite{Chekanov:2002xi} collaborations, and with the recent measurement by the ALICE collaboration~\cite{TheALICE:2014dwa} on ultraperipheral proton-lead collision (which can be seen as a photon-proton scattering due to the $Z^2$ enhancement for the photon flux emitted from the nucleus). The models are found to be in agreement with the current data\footnote{Compare with Ref.~\cite{Armesto:2014sma} where the IPnonsat model is not separately fitted to the HERA data, instead the same parameters are used in both parametrizations}, but the future more precise LHC data at even higher $W$ (requiring larger center-of-mass energy for the ultraperipheral proton-nucleus scattering or more forward rapidities) will be in the region where the difference between the models is large.

 \begin{figure}[tb]
\begin{center}
\includegraphics[width=0.49\textwidth]{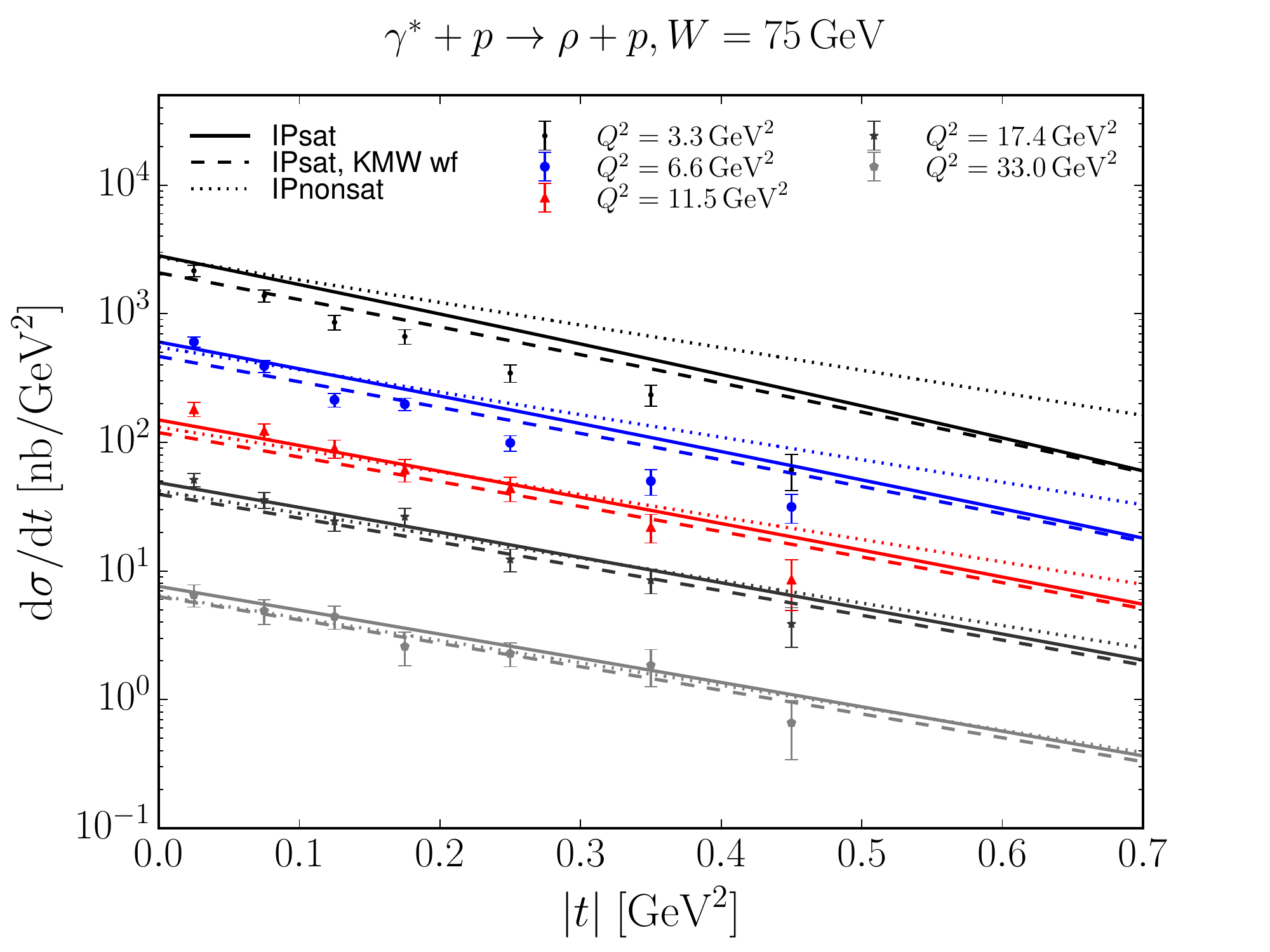}
\end{center}
\caption{
Differential $\rho$ electroproduction cross section as a function of momentum transfer $-t={\bf \Delta}^2$. The dashed line is obtained by using the $\rho$ wave function provided in Ref.~\cite{Kowalski:2006hc} where the light quark mass is $m_\text{light}=0.14\gev$. The other IPsat and IPnonsat curves use the wave function parametrizations from Tab.~\ref{table:boostedgaussian}. The experimental data are from the H1 collaboration~\cite{Aaron:2009xp}.  
}\label{fig:rhospectra}
\end{figure}

Let us then study the production of light mesons. The differential $\rho$ electroproduction cross section in different $Q^2$ bins is shown in Fig.~\ref{fig:rhospectra} and compared with the H1 data~\cite{Aaron:2009xp}. In the lowest $Q^2$ bins the applicability of our framework is questionable as, again, we do not have a large scale suppressing nonperturbative contributions. The agreement with the H1 data is good especially at higher $Q^2$ using both IPsat and IPnonsat parametrizations. Note that the light quark mass is much larger in the IPnonsat model, which explains why the cross section at small $-t$ is actually smaller in the IPnonsat calculation. Again, by calculating the $\rho$ production with the IPsat model parametrization and the KMW wave function~\cite{Kowalski:2006hc} we find that the larger light quark mass suppresses the cross section at small $Q^2$ bins similarly to the case of J/$\Psi$ production, and that the different impact parameter profile causes diffractive dips at large $|t|$ when we use the IPsat parametrization.

\section{Future experiments}
\label{Future}
\subsection{Proton targets}

\begin{figure}[tb]
\begin{center}
\includegraphics[width=0.49\textwidth]{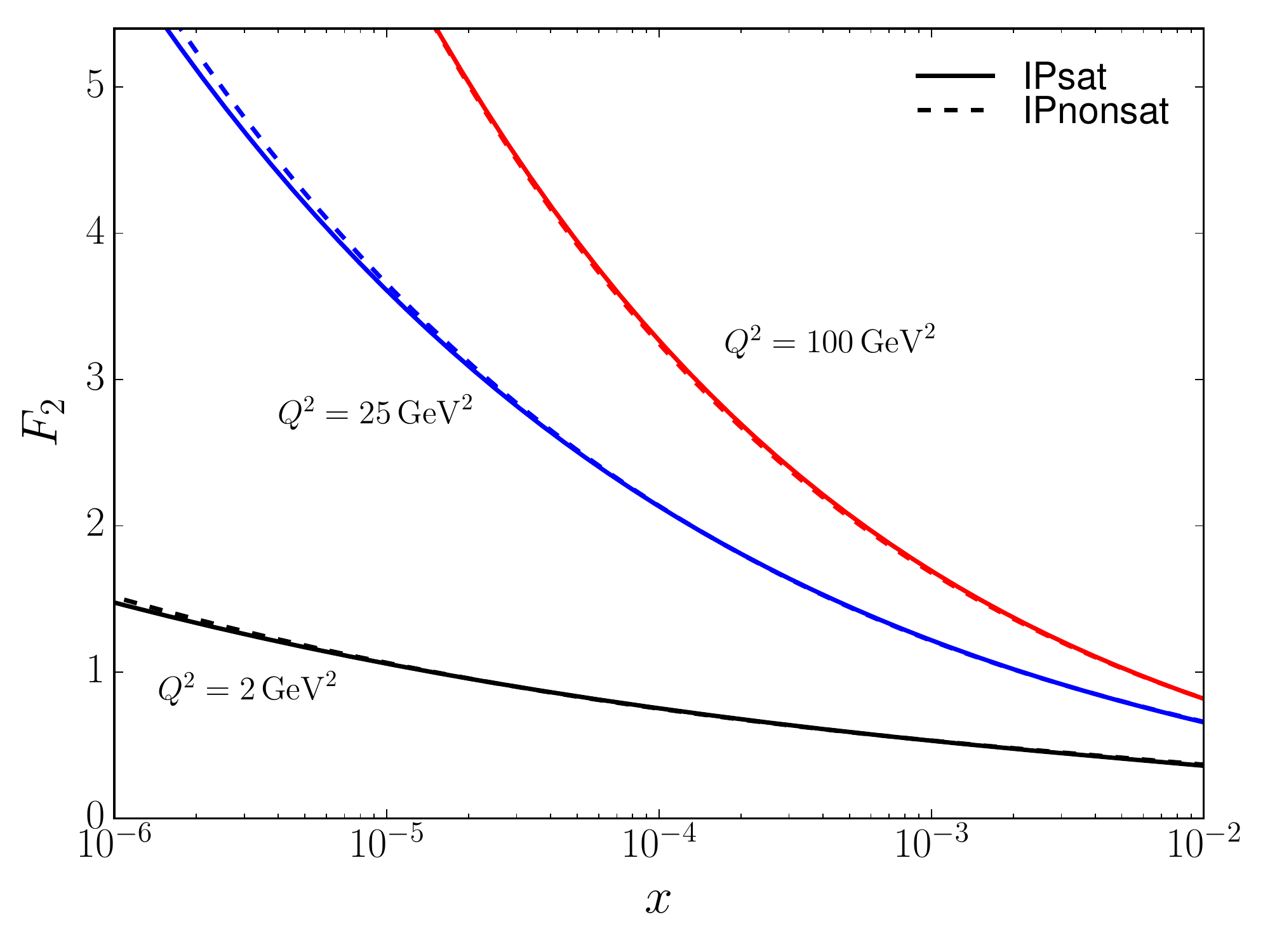}
\end{center}
\caption{Proton structure function $F_2$ computed using the IPsat and IPnonsat fits. For $F_L$, the difference between IPsat and IPnonsat parametrizations is similar. }
\label{fig:f2}
\end{figure}

As we saw in Sec. \ref{HERA}, both the IPsat and IPnonsat parametrizations give equally good descriptions of the HERA data. Due to the different behavior of the gluon distribution $xg$ at small $x$, and especially as the gluon distribution is eikonalized in the IPsat parametrization, differences are expected to arise when extrapolating to smaller values of Bjorken-$x$. This we already found in the case of exclusive J/$\Psi$ production in Fig.~\ref{fig:jpsi_totxs}, where both parametrizations give comparable results in the range covered by the HERA data, but differ by $\sim 50\%$ in the kinematics covered by recent and near-future LHC experiments. On the contrary, for inclusive DIS the structure function $F_2$ predicted by both, the IPsat and IPnonsat, prametrizations overlap almost perfectly for a wide range of $x$ values, as shown in Fig.~\ref{fig:f2}. Even at very small $x\sim 10^{-6}$ the two models differ only at the level of a few percent. 

\begin{figure}[tb]
\begin{center}
\includegraphics[width=0.49\textwidth]{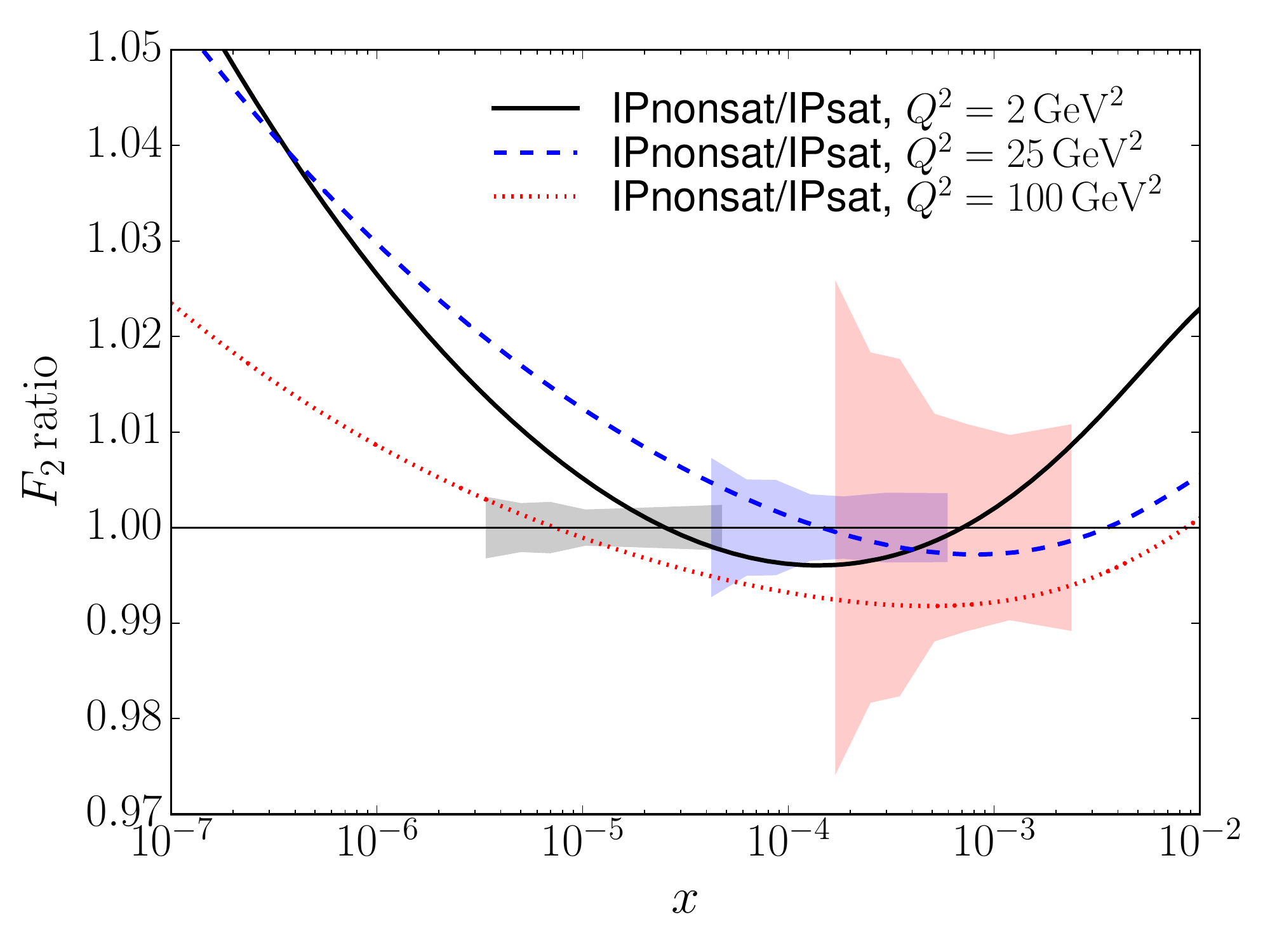}
\end{center}
\caption{
Ratio of the structure function $F_2$ computed using the IPsat and IPnonsat fits. The bands show relative uncertainty projected for the LHeC assuming the IPsat parametrization from small (left) to large (right) $Q^2$ (see text).}
\label{fig:f2ratio}
\end{figure}

To see if the future high-energy DIS experiments can measure the structure functions with an accuracy lower or comparable to the difference between the two models, we show in Fig.~\ref{fig:f2ratio} the ratio of $F_2$ obtained using the IPnonsat and IPsat parametrizations compared with the projected accuracy of the LHeC measurement~\cite{AbelleiraFernandez:2012cc}. 
As the uncertainty estimates for the LHeC consist of projected absolute, not relative uncertainties, the relative uncertainties shown as a colorfull bands are obtained by comparing the projected uncertainty to the result obtained by applying the IPsat parametrization. As already seen in Fig.~\ref{fig:f2}, the differences are at a few percent level, and only slightly larger than the projected experimental accuracy at the LHeC. FCC-eh would probe $x$ values down to $x\sim 10^{-7}$ with comparable precision, thus at least in $F_2$ there would not be a striking difference between the IPsat and IPnonsat extrapolations. For $F_L$ the model differences are similar, but the experimental accuracy is much lower. 

The data from future DIS machines on $F_2$, $F_{2,\text{charm}}$ and $F_L$ will thus make it possible to constrain dipole-proton scattering much more accurately, thanks to the fact that different observables are sensitive to different dipole sizes. However, in inclusive $e+p$ scattering the IPsat model predicts the non-linear effects to be small. It thus becomes necessary to study nuclear DIS, where one expects the saturation effects to be enhanced by a large factor $A^{1/3}$.

\subsection{Nuclear targets}
\label{sec:nuclei}

Using the Optical Glauber model one can extend the dipole-proton scattering amplitude to the dipole-nucleus one. 
Calculating the dipole-nucleus scattering by averaging over the positions of the individual nucleons from the Woods Saxon distribution following Ref.~\cite{Kowalski:2003hm} one obtains
\begin{equation}
\label{eq:ipsatlumpynuke}
\frac {\ud \sigmaa}{\ud^2 \bt} = 2\left[1 - \left( 1 - \frac{1}{2} T_A(\bt) \sigmadip \right)^A \right],
\end{equation}
where $\sigmadip$ is the total dipole-proton cross section integrated over impact parameter, see Eq.~\eqref{eq:ipsat}. For large nuclei, this gives
\begin{equation}
\label{eq:ipsatlumpynuke_largeA}
\frac {\ud \sigmaa}{\ud^2 \bt} = 2\left[1 - \exp \left( 1 - \frac{1}{2} AT_A(\bt) \sigmadip \right) \right].
\end{equation}
Only if, in addition to having a large $A$, the dipole-proton cross section is small (which requires small $r$ as $\sigmadip \sim \ln r$) one obtains the \emph{smooth nucleus} result
\begin{equation}
\label{eq:ipsatnuke}
\frac {\ud \sigmaa}{\ud^2 \bt} = 2\left[1 - \exp \left( -r^2 F(x,r) A T_A(\bt) \right) \right].
\end{equation}
In practice, as large dipoles have numerically significant contribution to $F_2$, this approximation is not a realistic and results in too small nuclear suppression as discussed in Ref.~\cite{Kowalski:2003hm}. 
Here $T_A$ is the Woods Saxon distribution integrated over the longitudinal coordinate, and the nuclear radius is $R_A=1.13A^{1/3} - 0.86 A^{-1/3}$ fm. The normalization is chosen such that $\int \der^2 \bt T_A(\bt)=1$. The corresponding dipole-nucleus amplitude in the IPnonsat model is the first term from the series expansion
\begin{equation}
\label{eq:ipnonsatnuke}
\frac {\ud \sigmaa}{\ud^2 \bt} =  2 r^2 F(x,r) A T_A(\bt) .
\end{equation}

\begin{figure}[tb]
\begin{center}
\includegraphics[width=0.49\textwidth]{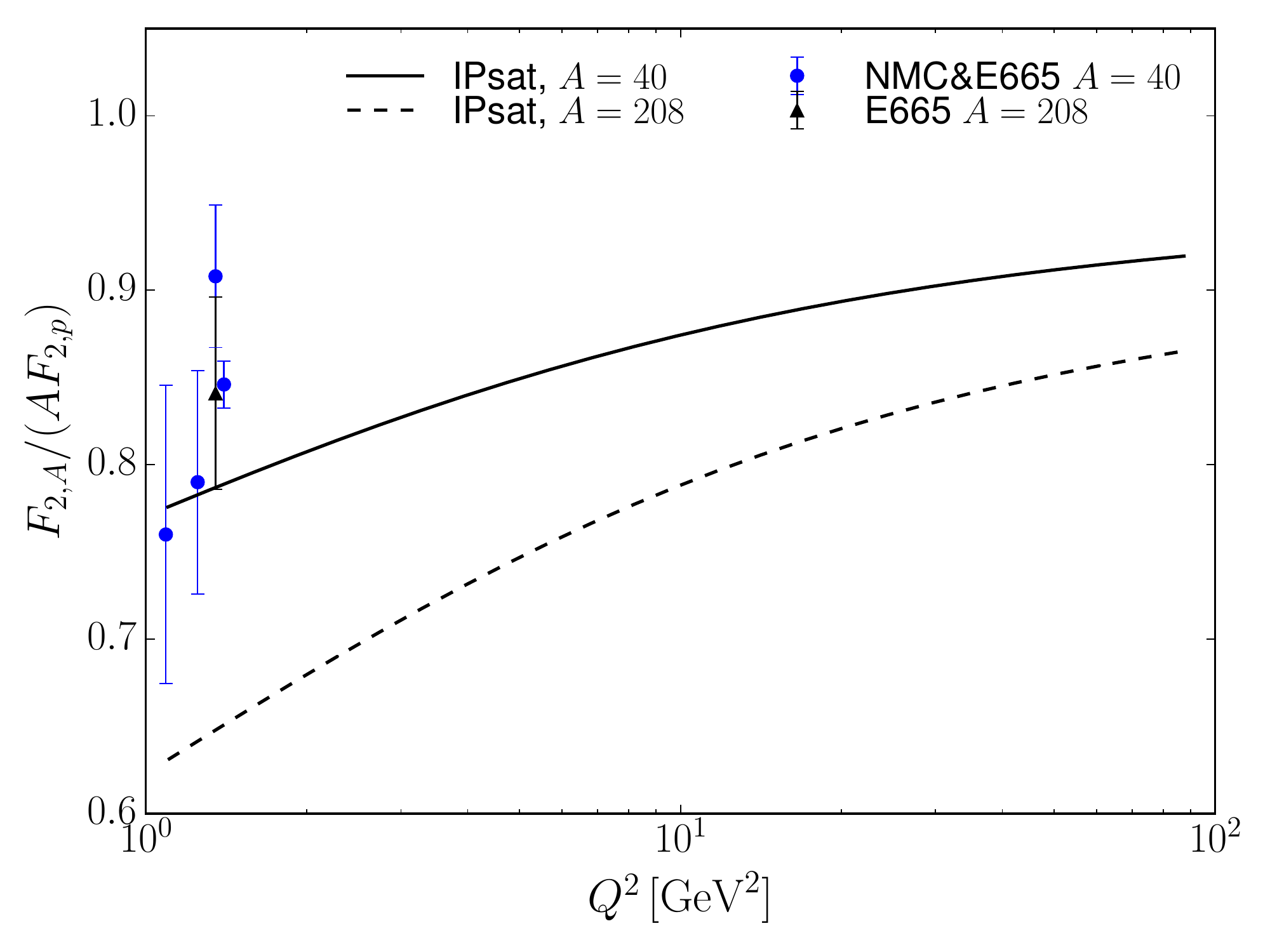}
\end{center}
\caption{Nuclear suppression for the structure function $F_2$ compared with the NMC and E665 data~\cite{Amaudruz:1995tq,Adams:1995is,Arneodo:1995cs}. The lead results are at $x=0.006185$. The Calcium data points cover $x$ values $0.005 \dots 0.0085$, and our calculation is done at average $x=0.0068$. By construction this ratio is exactly 1 with the IPnonsat parametrization. }

\label{fig:suppression_f2_exp}
\end{figure}

The nuclear suppression factor for the structure function $F_2$ is shown in Fig.~\ref{fig:suppression_f2_exp}, where we calculate
\begin{equation}
	R=\frac{F_{2,A}}{AF_{2,p}}.
\end{equation}
For comparison, the experimental data points for Calcium and Lead from \cite{Amaudruz:1995tq,Adams:1995is,Adams:1995is} are shown\footnote{Part of the fixed target data is obtained by comparing total cross sections for nuclear and deuteron targets, which is not exactly the same as our structure function ratio}.

\begin{figure}[tb]
\begin{center}
\includegraphics[width=0.49\textwidth]{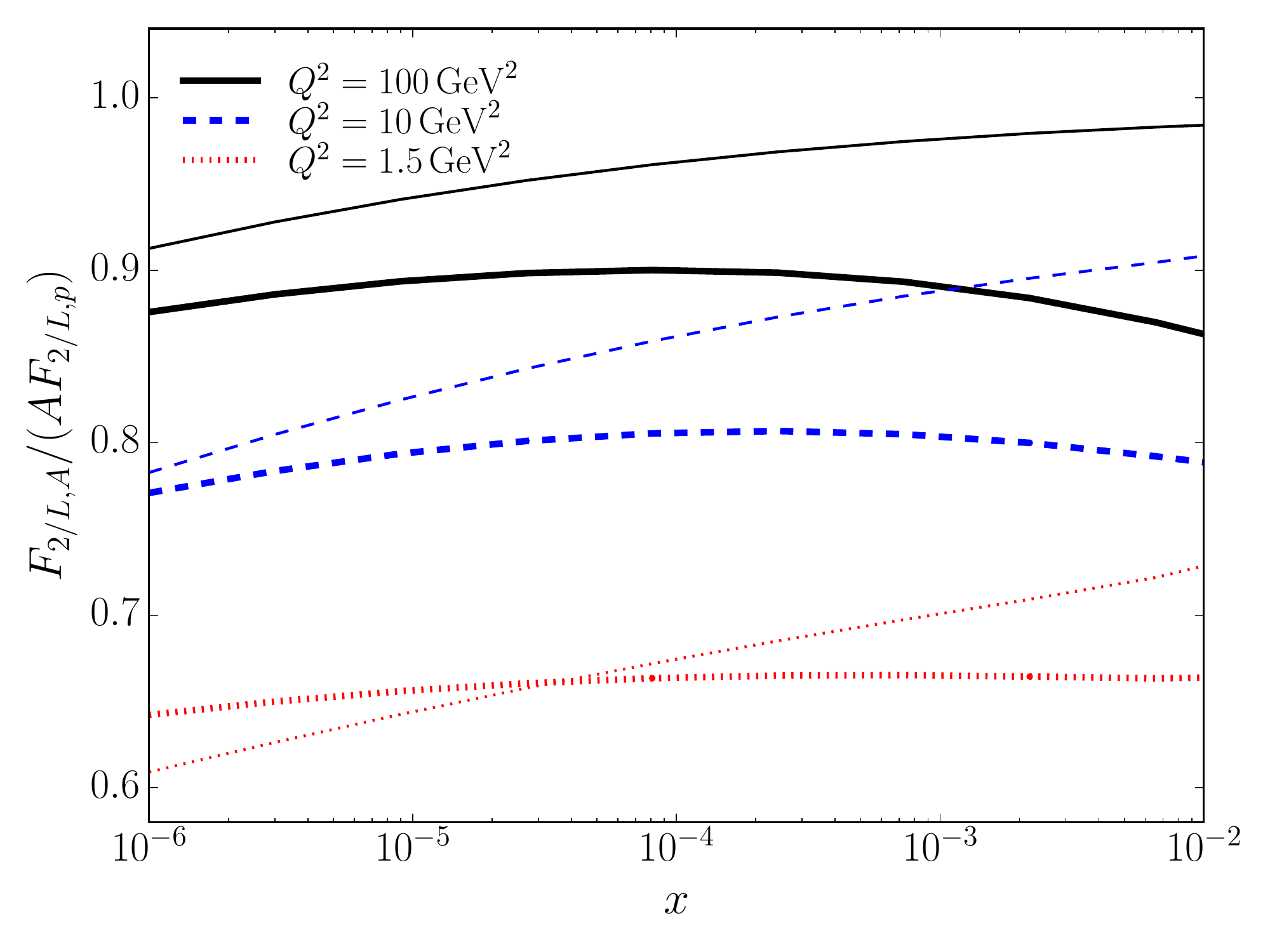}
\end{center}
\caption{Bjorken-$x$ dependence of the nuclear suppression factor for $F_2$ (thick lines) and $F_L$ (thin lines). By construction this ratio is exactly 1 with the IPnonsat parametrization.  }
\label{fig:suppression_f2fl_x}
\end{figure}

The Bjorken-$x$ dependence of the $F_2$ and $F_L$ nuclear suppression factors is studied in Fig.~\ref{fig:suppression_f2fl_x}. We find that in the case of $F_2$, the $x$ dependence is weak due to the fact that a significant part of $F_2$ originates from large saturated dipoles, whose scattering cross section is not affected by the relatively slow evolution of the saturation scale $Q_s^2 \sim xg(x,\mu^2)$. In the case of $F_L$, which is dominated by smaller dipoles, a significantly faster $x$ dependence is found.

We also find that when $x$ increases, at some point the nuclear suppression starts to \emph{increase}, which is counter intuitive. 
A similar observation was already seen in Ref.~\cite{Kowalski:2007rw}. We note that if we were to do a BK evolution for the nucleus similarly as in Ref.~\cite{Lappi:2013zma}, we would get a suppression factor for $F_2$ that always increases with increasing $x$. We consider the fact that the $F_2$ suppression factor has a maximum as a function of $x$ to be an artefact of the shortcomings of the IPsat parametrization (e.g. decreasing $xg$ with increasing scale in Fig.~\ref{fig:xg} which effectively decreases the saturation scale probed by larger dipoles).

Let us next study nuclear suppression in exclusive vector meson production. Here, we analyze the $Q^2$ dependence of light $\rho$ and $\phi$ meson production that allows us to scan the transition from saturation to dilute region (see also Ref.~\cite{Mantysaari:2017slo}). In addition, we include $J/\Psi$, which is significantly smaller and heavier, and should experience less non-linear effects. As the total coherent cross section scales like $A^2$, and the width of the first diffractive peak is proportional to $1/R_A^2 \sim A^{-2/3}$, we study the 
suppression factor
\begin{equation}
R=\frac{\sigma^{\gamma^* A \to V A}} {c A^{4/3} \sigma^{\gamma p \to Vp}},
\end{equation}
where $V$ refers to the vector meson species. Note that diffractive cross sections are enhanced more strongly by the large nucleus compared to the inclusive scattering which scales linearly in $A$. The numerical factor $c$ in the denominator can be obtained as a ratio of the form factors for the nucleus and the proton 
\begin{equation}
c = \frac{A^2 \int \der t \left| \tilde T_A(\sqrt{t}) \right|^2 } {A^{4/3} \int \der t \left | \tilde T_p(\sqrt{t}) \right|^2 },
\end{equation}
as the form factors determine the $t$ spectra. 
Here the form factors are $\tilde T_A(\sqrt{t}) = \int \der^2 \bt e^{-i \bt \cdot \Deltat} T_A(\bt)$ and $\tilde T_p(\sqrt{t})=\int \der^2 \bt e^{-i \bt \cdot \Deltat} T_p(\bt)$ with $t=\Deltat^2$. For the Gold nucleus with $A=197$, this gives $c\approx 0.5011$. By construction, this definition gives $R=1$ in case of the linear IPnonsat parametrization.

\begin{figure}[tb]
\begin{center}
\includegraphics[width=0.49\textwidth]{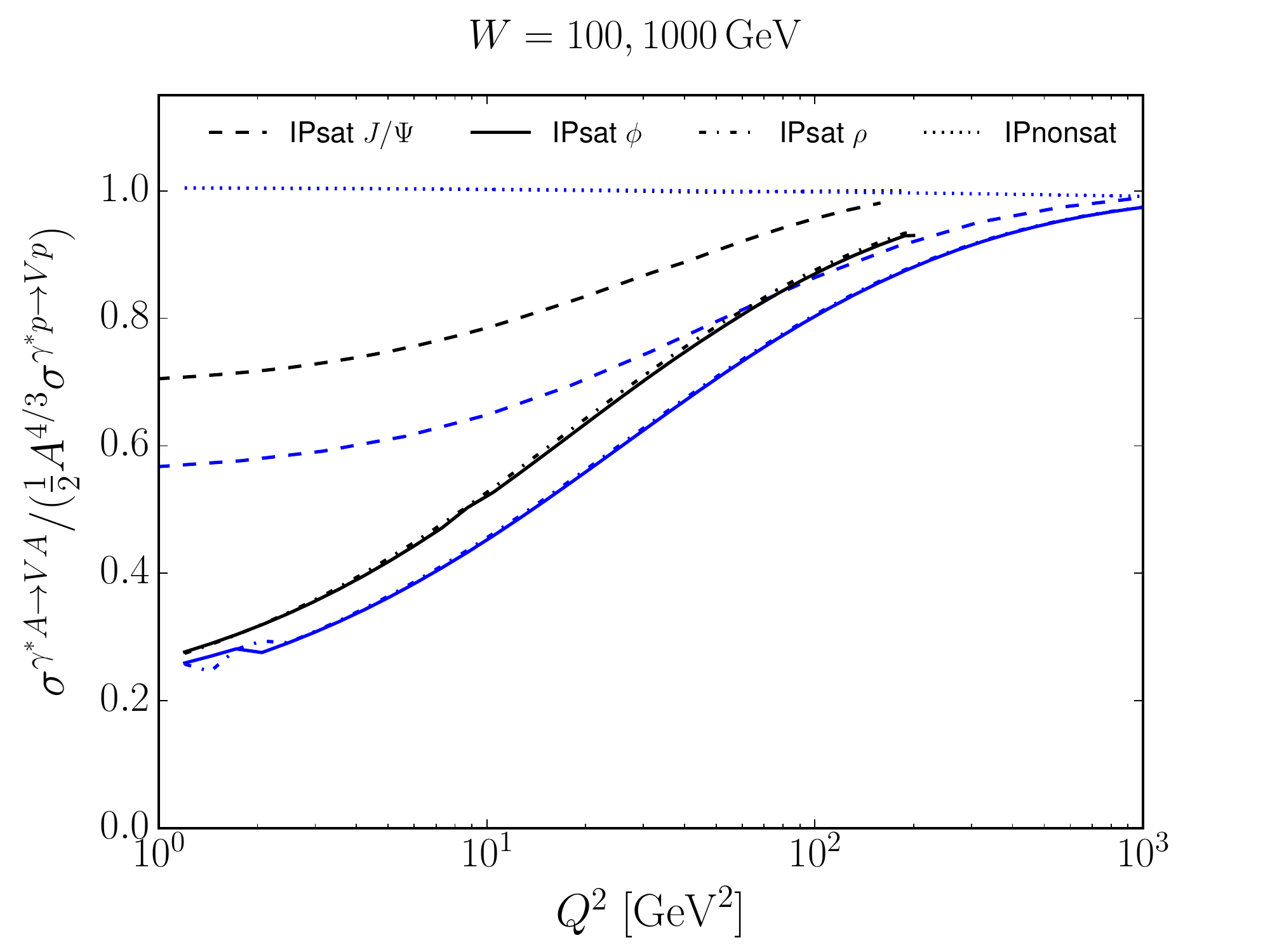}
\end{center}
\caption{Nuclear suppression for total coherent J/$\Psi$, $\rho$ and $\phi$ vector meson production as a function of $Q^2$. The upper black lines refer to the case where $W=100\gev$ and the lower blue lines are calculated at $W=1000\gev$. The results are shown in the range where $\xpom < 0.02$.
}
\label{fig:suppression_rho_phi}
\end{figure}

The $Q^2$ dependence of the suppression factor is relatively weak as shown in Fig.~\ref{fig:suppression_rho_phi}, with $R$ reaching unity only at $Q^2 \sim 1000\gev^2$. Physically the reason here is that even if at large $Q^2$ the photon preferably splits into a small dipole which does not see nonlinear effects, the requirement that a light (and large) vector meson is formed at the final state gives a small weight for the small dipoles. Instead, the specified final state requires the dipole to be relatively large, and it becomes necessary to go to very large $Q^2$ to give enough weight on small dipoles so that the suppression factor becomes close to 1. In case of J/$\Psi$ the vector meson wave function always picks up relatively small dipoles, thus the maximum suppression is only around $0.7$. For a discussion about the nuclear suppression in incoherent scattering, the reader is referred to Ref.~\cite{Lappi:2010dd}.

In Fig.~\ref{fig:suppression_rho_phi} we also show the $W$ dependence of the suppression factor, which is found to be relatively modest. However, in the future Electron Ion Collider it will be useful to have maximally large $Q^2$ lever arm to study the evolution of the suppression factor from the saturated to the dilute region (see also Refs.~\cite{Mantysaari:2017slo,Aschenauer:2017jsk}). For example, in the case of $\rho$ production and at $\xpom=10^{-2}$ (which is around the maximum $x$ where our model can be considered to be valid), the maximum $Q^2$ that can be reached at an EIC with $\sqrt{s_{NN}}=90\gev$ is $Q^2_\text{max}\approx 100\gev^2$, which would make it possible to observe the evolution of the nuclear suppression from $R\sim 0.2$ to $R\sim 0.8$.

\section{Conclusions}
\label{conclusions}

The possibility of finding new and exciting QCD phenomena is just around the corner, with the next generation of DIS colliders to come soon. In view of this it is timely to exploit to the fullest the available data. In this path this work bring us one step forward, by determining not only the IPsat model with the modern data sets but also for the first time the linearized IPnonsat parametrization is determined from the same data in a fully consistent way. 

The main differences to previous works are the inclusion of the charm data to the global fit, which allows us to constrain the quark masses, and the use of a variable flavor number scheme. Also, for the first time, the combined HERA I+II datasets are used in dipole model fits with a similar outcome than in case of the HERA I combined results. We find that both models, with and without saturation, result in almost identical cross sections at HERA kinematics, and that the differences in 
$e+p$ scattering are expected to be small even in the LHeC or FCC-eh kinematics. The nonlinear effects, however, become significant if a nuclear target/beam is used and should be easily observed in the future Electron-Ion Collider. 

Despite some differences in the setup with the previous literature, the resulting dipole amplitude and calculated cross sections are similar than in the previous work~\cite{Rezaeian:2012ji}. This is a consequence of performing the fits using comparable data sets, which extrapolates to similar dipole amplitudes. Therefore for the IPsat case the models found in the literature will provide reasonable numerical results in the kinematical range accessible in current and future colliders. We emphasize that having a linearized ``IPnonsat'' model independently constrained by the HERA data is necessary for estimating the size of the saturation effects in these experiments.

The similar cross sections obtained from both IPsat and IPnonsat parametrizations are understood in terms of the effective description of the confinement scale physics. The linearized dipole cross section violates unitarity at large dipoles, and the fit compensates that by imposing an effective confinement effect damping dipoles larger than $\sim 1/m_{\text{light}}$. In the IPsat model, the unitarity requirement limits the contribution from unphysically large dipoles and a large light quark mass is not required in order to obtain a good description of the HERA data. 

The inclusive structure function $F_2$ (and thus the reduced cross section $\sigma_r$) is especially sensitive to the dipoles expected to be heavily influenced by confinement effects not completely included in this work. On the other hand, $F_L$ and $F_{2,\text{charm}}$ are not sensitive at all to dipoles larger than $\sim 1/\lqcd$. Thus, the future more precise $F_L$ and $F_{2,\text{charm}}$ data, together with inclusive structure function measurements, will allow us to perform a much more precise test of the saturation picture.

Both IPsat and IPnonsat parametrizations give comparable predictions for structure functions at the energies available in future DIS experiments such as LHeC and FCC-eh. Slightly larger differences are seen when calculating predictions for exclusive vector meson production, but in order to really see the onset of non-linear nature of QCD, we find that it is crucial to perform DIS with nuclear targets. These effects should become clearly visible already at the EIC energies. Additionally, the potential for inclusive diffraction to separate between the linear and non-linear parametrizations could be studied in  future work.

 \section*{Acknowledgements}
We thank T. Lappi, T. Toll, T. Ullrich and R. Venugopalan for discussions and are grateful to Max Klein for providing LHeC and FCC-eh pseudodata for the present work. H. M. was supported under DOE Contract No. DE-SC0012704 and European Research Council, Grant ERC-2015-CoG-681707, and wishes to thank the Nuclear Theory Group at BNL for hospitality during the preparation of this manuscript. P. Z. acknowledges the support by the U.S. Department of Energy under contract number No. DE-SC0012704.

\appendix

\section{Contribution from the bottom quark}
\label{appendix:bquark}
Recently the H1 and ZEUS collaborations have released the first combined bottom quark contribution to the reduced cross section~\cite{H1:2018flt}. Due to the relatively large uncertainties and limited number of datapoints, we did not include this unpublished dataset in our fit. Instead, we can use it to check that the bottom quark contribution included in the calculation of the inclusive cross section is compatible with the current measurements.

\begin{figure*}[tb]
\begin{center}
\includegraphics[width=0.7 \textwidth]{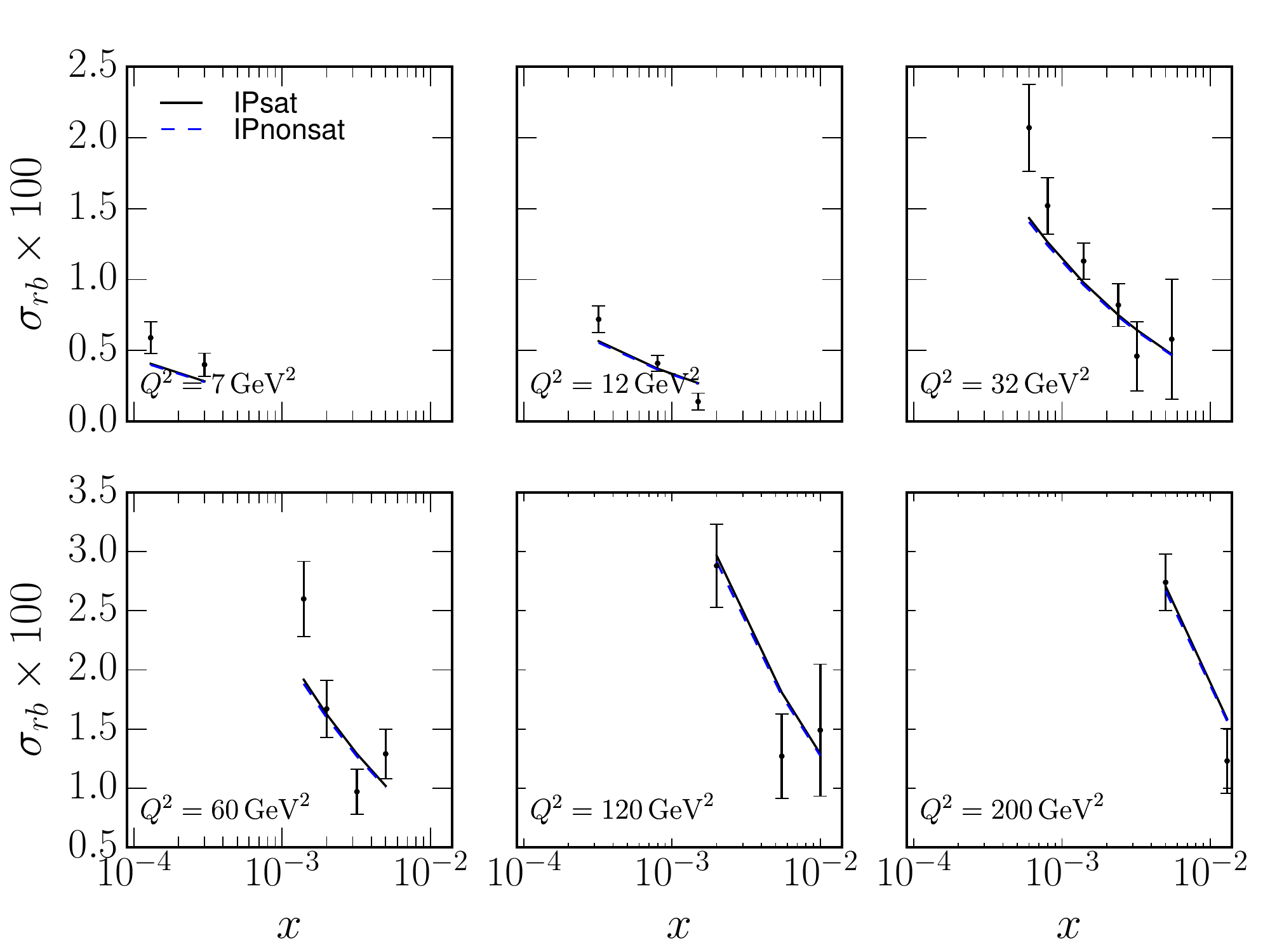}
\end{center}
\caption{Contribution to the reduced cross section from $b$ quarks, compared with the combined HERA data~\cite{H1:2018flt}. }
\label{fig:sigmar_b}
\end{figure*}

The predicted bottom quark reduced cross section compared with the HERA data is shown in Fig.~\ref{fig:sigmar_b}. As noted for the inclusive and charm reduced cross sections, the IPsat and IPnonsat parametrizations give approximately equal results in the HERA kinematics. The description of the data is also good, the $\chi^2/N$ being $1.81$ ($1.90$) for the IPsat (IPnonsat) model, when comparing with datapoints at $Q^2\le 500\gev^2$. Slightly faster $Q^2$ evolution is obtained compared with the HERA measurements, similarly as in case of charm reduced cross section (see Fig.~\ref{fig:charm_sigmar}).

\section{Effect of the HERA I+II data}
\label{appendix:hera_i_ii}
\begin{figure*}[tb]
\centering
\begin{minipage}{0.45\textwidth}
\centering
	\includegraphics[width=\textwidth]{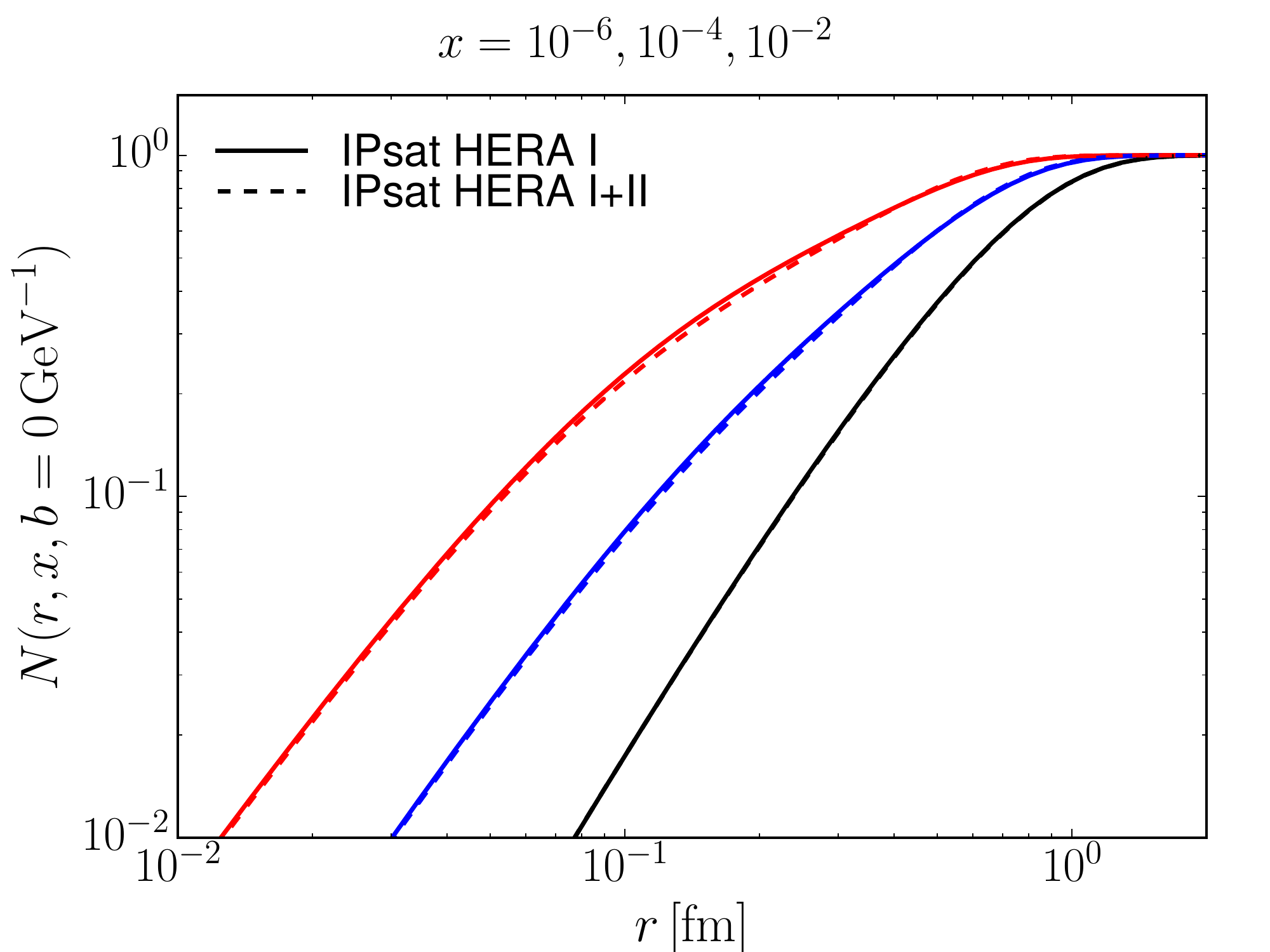}
\caption{Comparison of the IPsat dipole amplitudes obtained by fits to HERA I and HERA I+II data. The used parametrizations are the first and second line of Table~\ref{tab:fits}. Bjorken $x$ values decrease to the left.
}
\label{fig:dipole_hera_i_ii}
\end{minipage} \quad 
\begin{minipage}{0.45\textwidth}
\centering
\includegraphics[width=\textwidth]{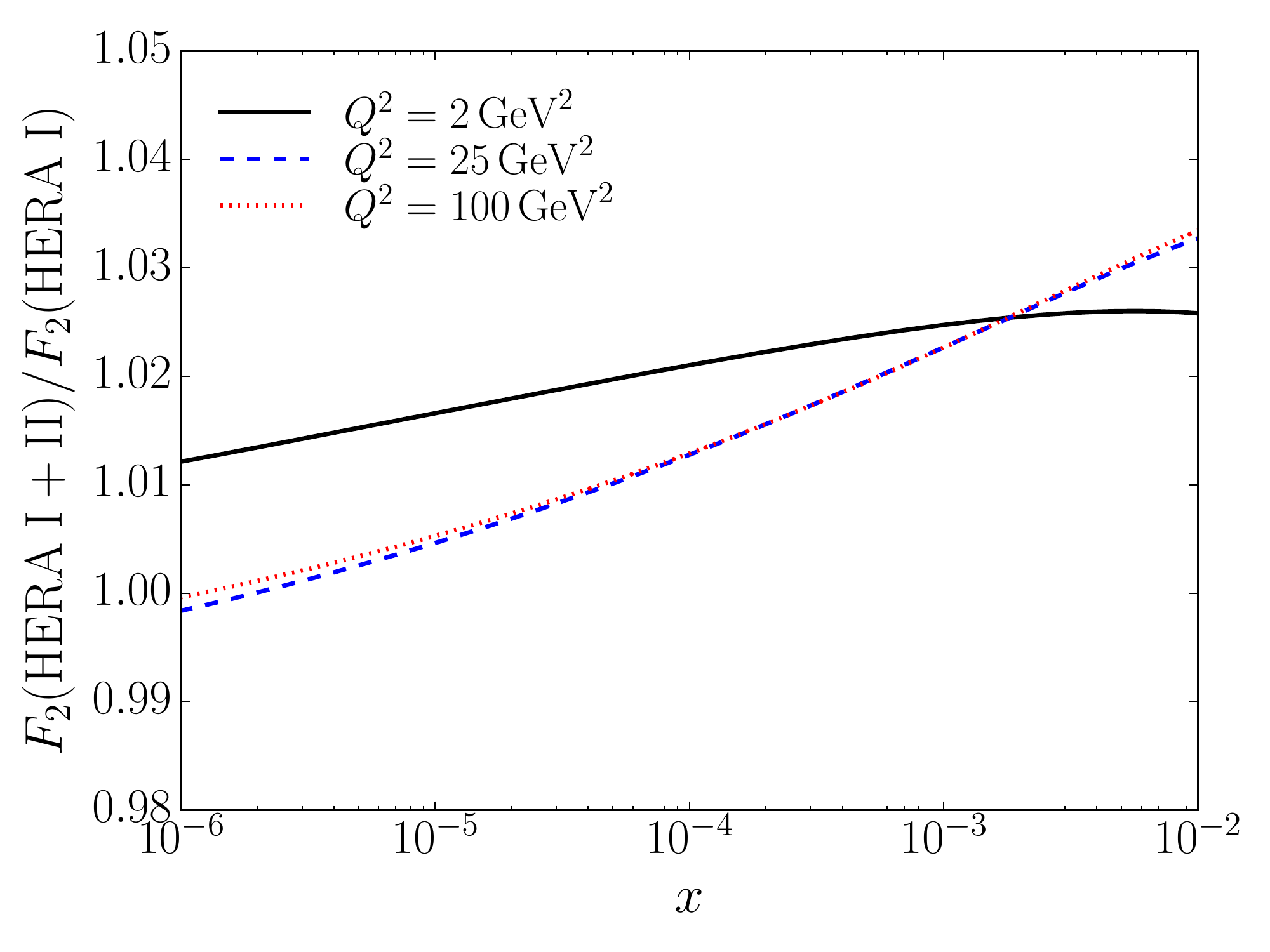}
\caption{Ratio of the $F_2$ structure functions obtained by fits to HERA I+II and HERA I data using the IPsat model. The used parametrizations are the first and second line of Table~\ref{tab:fits}. 
}
\label{fig:f2_hera_i_ii}
\end{minipage}
\end{figure*}

In this work we have considered the fit performed to HERA I data to be the main result of this work. As was shown in Table~\ref{tab:fits}, the best fit parameters change slightly when one fits the final HERA I+II dataset which, in addition to having smaller uncertainties, is more dominated by the inclusive reduced cross section. In order to quantify the effect of different datasets on the fit result, we show in Fig.~\ref{fig:dipole_hera_i_ii}. The obtained dipole amplitudes are found to be very similar over a broad range in $x$. Consequently, the structure function $F_2$ obtained with both parametrizations is practically identical. This is demonstrated by showing in Fig.~\ref{fig:f2_hera_i_ii} the $F_2$ obtained using the fit result to HERA I+II data (second line in Table~\ref{tab:fits}) normalized by the result obtained by applying the HERA I fit result (first line in Table~\ref{tab:fits}). Similar results are found in case of $F_L$. Thus, both fits results can be considered to be equivalent .

\bibliographystyle{JHEP-2modlong.bst}
\bibliography{../../refs}

\providecommand{\href}[2]{#2}\begingroup\raggedright\begin{thebibliography}{10}

\bibitem{Gribov:1972ri}
V.~N. Gribov and L.~N. Lipatov, {\it {Deep inelastic e p scattering in
  perturbation theory}},  {\em Sov. J. Nucl. Phys.} {\bf 15} (1972) 438.
\newblock [Yad. Fiz.15,781(1972)].

\bibitem{Gribov:1972rt}
V.~N. Gribov and L.~N. Lipatov, {\it {$e^+$ $e^-$ pair annihilation and deep
  inelastic e p scattering in perturbation theory}},  {\em Sov. J. Nucl. Phys.}
  {\bf 15} (1972) 675.
\newblock [Yad. Fiz.15,1218(1972)].

\bibitem{Altarelli:1977zs}
G.~Altarelli and G.~Parisi, {\it {Asymptotic Freedom in Parton Language}},
  \href{http://dx.doi.org/10.1016/0550-3213(77)90384-4}{{\em Nucl. Phys.} {\bf
  B126} (1977) 298}.

\bibitem{Dokshitzer:1977sg}
Y.~L. Dokshitzer, {\it {Calculation of the Structure Functions for Deep
  Inelastic Scattering and $e^+$ $e^-$ Annihilation by Perturbation Theory in
  Quantum Chromodynamics.}},  {\em Sov. Phys. JETP} {\bf 46} (1977) 641.

\bibitem{Gelis:2010nm}
F.~Gelis, E.~Iancu, J.~Jalilian-Marian and R.~Venugopalan, {\it {The Color
  Glass Condensate}},
  \href{http://dx.doi.org/10.1146/annurev.nucl.010909.083629}{{\em Ann. Rev.
  Nucl. Part. Sci.} {\bf 60} (2010) 463}
  [\href{http://arXiv.org/abs/1002.0333}{{\tt arXiv:1002.0333 [hep-ph]}}].

\bibitem{Kowalski:2003hm}
H.~Kowalski and D.~Teaney, {\it {An Impact parameter dipole saturation model}},
   \href{http://dx.doi.org/10.1103/PhysRevD.68.114005}{{\em Phys. Rev.} {\bf
  D68} (2003) 114005} [\href{http://arXiv.org/abs/hep-ph/0304189}{{\tt
  arXiv:hep-ph/0304189 [hep-ph]}}].

\bibitem{Kowalski:2006hc}
H.~Kowalski, L.~Motyka and G.~Watt, {\it {Exclusive diffractive processes at
  HERA within the dipole picture}},
  \href{http://dx.doi.org/10.1103/PhysRevD.74.074016}{{\em Phys. Rev.} {\bf
  D74} (2006) 074016} [\href{http://arXiv.org/abs/hep-ph/0606272}{{\tt
  arXiv:hep-ph/0606272}}].

\bibitem{Rezaeian:2012ji}
A.~H. Rezaeian, M.~Siddikov, M.~Van~de Klundert and R.~Venugopalan, {\it
  {Analysis of combined HERA data in the Impact-Parameter dependent Saturation
  model}},  \href{http://dx.doi.org/10.1103/PhysRevD.87.034002}{{\em Phys.
  Rev.} {\bf D87} (2013) 034002} [\href{http://arXiv.org/abs/1212.2974}{{\tt
  arXiv:1212.2974}}].

\bibitem{Luszczak:2013rxa}
A.~Luszczak and H.~Kowalski, {\it {Dipole model analysis of high precision HERA
  data}},  \href{http://dx.doi.org/10.1103/PhysRevD.89.074051}{{\em Phys. Rev.}
  {\bf D89} (2014)~no.~7 074051} [\href{http://arXiv.org/abs/1312.4060}{{\tt
  arXiv:1312.4060 [hep-ph]}}].

\bibitem{Luszczak:2016bxd}
A.~Luszczak and H.~Kowalski, {\it {Dipole model analysis of highest precision
  HERA data, including very low $Q^{2}$'s}},
  \href{http://dx.doi.org/10.1103/PhysRevD.95.014030}{{\em Phys. Rev.} {\bf
  D95} (2017)~no.~1 014030} [\href{http://arXiv.org/abs/1611.10100}{{\tt
  arXiv:1611.10100 [hep-ph]}}].

\bibitem{Aaron:2009aa}
{\bf H1 and ZEUS} collaborations, F.~Aaron {\em et.~al.}, {\it {Combined
  Measurement and QCD Analysis of the Inclusive $e^\pm p$ Scattering Cross
  Sections at HERA}},  \href{http://dx.doi.org/10.1007/JHEP01(2010)109}{{\em
  JHEP} {\bf 1001} (2010) 109} [\href{http://arXiv.org/abs/0911.0884}{{\tt
  arXiv:0911.0884 [hep-ex]}}].

\bibitem{Abramowicz:1900rp}
{\bf H1 and ZEUS} collaborations, H.~Abramowicz {\em et.~al.}, {\it {Combination
  and QCD Analysis of Charm Production Cross Section Measurements in
  Deep-Inelastic ep Scattering at HERA}},
  \href{http://dx.doi.org/10.1140/epjc/s10052-013-2311-3}{{\em Eur. Phys. J.}
  {\bf C73} (2013) 2311} [\href{http://arXiv.org/abs/1211.1182}{{\tt
  arXiv:1211.1182 [hep-ex]}}].

\bibitem{Abramowicz:2015mha}
{\bf H1 and ZEUS} collaborations, H.~Abramowicz {\em et.~al.}, {\it {Combination
  of measurements of inclusive deep inelastic ${e^{\pm }p}$ scattering cross
  sections and QCD analysis of HERA data}},
  \href{http://dx.doi.org/10.1140/epjc/s10052-015-3710-4}{{\em Eur. Phys. J.}
  {\bf C75} (2015) 580} [\href{http://arXiv.org/abs/1506.06042}{{\tt
  arXiv:1506.06042 [hep-ex]}}].

\bibitem{H1:2018flt}
{\bf H1 and ZEUS} collaborations, H.~Abramowicz {\em et.~al.}, {\it {Combination
  and QCD analysis of charm and beauty production cross-section measurements in
  deep inelastic $ep$ scattering at HERA}},
  \href{http://dx.doi.org/10.1140/epjc/s10052-018-5848-3}{{\em Eur. Phys. J.}
  {\bf C78} (2018)~no.~6 473} [\href{http://arXiv.org/abs/1804.01019}{{\tt
  arXiv:1804.01019 [hep-ex]}}].

\bibitem{Ball:2017otu}
R.~D. Ball, V.~Bertone, M.~Bonvini, S.~Marzani, J.~Rojo and L.~Rottoli, {\it
  {Parton distributions with small-x resummation: evidence for BFKL dynamics in
  HERA data}},  \href{http://arXiv.org/abs/1710.05935}{{\tt arXiv:1710.05935
  [hep-ph]}}.

\bibitem{GolecBiernat:1998js}
K.~J. Golec-Biernat and M.~Wusthoff, {\it {Saturation effects in deep inelastic
  scattering at low $Q^2$ and its implications on diffraction}},
  \href{http://dx.doi.org/10.1103/PhysRevD.59.014017}{{\em Phys. Rev.} {\bf
  D59} (1998) 014017} [\href{http://arXiv.org/abs/hep-ph/9807513}{{\tt
  arXiv:hep-ph/9807513 [hep-ph]}}].

\bibitem{GolecBiernat:1999qd}
K.~J. Golec-Biernat and M.~Wusthoff, {\it {Saturation in diffractive deep
  inelastic scattering}},
  \href{http://dx.doi.org/10.1103/PhysRevD.60.114023}{{\em Phys. Rev.} {\bf
  D60} (1999) 114023} [\href{http://arXiv.org/abs/hep-ph/9903358}{{\tt
  arXiv:hep-ph/9903358 [hep-ph]}}].

\bibitem{Kowalski:2008sa}
H.~Kowalski, T.~Lappi, C.~Marquet and R.~Venugopalan, {\it {Nuclear enhancement
  and suppression of diffractive structure functions at high energies}},
  \href{http://dx.doi.org/10.1103/PhysRevC.78.045201}{{\em Phys. Rev.} {\bf
  C78} (2008) 045201} [\href{http://arXiv.org/abs/0805.4071}{{\tt
  arXiv:0805.4071 [hep-ph]}}].

\bibitem{Accardi:2012qut}
A.~Accardi {\em et.~al.}, {\it {Electron Ion Collider: The Next QCD Frontier -
  Understanding the glue that binds us all}},
  \href{http://dx.doi.org/10.1140/epja/i2016-16268-9}{{\em Eur. Phys. J.} {\bf
  A52} (2016) 268} [\href{http://arXiv.org/abs/1212.1701}{{\tt arXiv:1212.1701
  [nucl-ex]}}].

\bibitem{Aschenauer:2017jsk}
E.~C. Aschenauer, S.~Fazio, J.~H. Lee, H.~Mäntysaari, B.~S. Page, B.~Schenke,
  T.~Ullrich, R.~Venugopalan and P.~Zurita, {\it {The Electron-Ion Collider:
  Assessing the Energy Dependence of Key Measurements}},
  \href{http://arXiv.org/abs/1708.01527}{{\tt arXiv:1708.01527 [nucl-ex]}}.

\bibitem{AbelleiraFernandez:2012cc}
{\bf LHeC Study Group} collaboration, J.~Abelleira~Fernandez {\em et.~al.},
  {\it {A Large Hadron Electron Collider at CERN: Report on the Physics and
  Design Concepts for Machine and Detector}},
  \href{http://dx.doi.org/10.1088/0954-3899/39/7/075001}{{\em J. Phys.} {\bf
  G39} (2012) 075001} [\href{http://arXiv.org/abs/1206.2913}{{\tt
  arXiv:1206.2913 [physics.acc-ph]}}].

\bibitem{Zimmermann:2014qxa}
F.~Zimmermann, M.~Benedikt, D.~Schulte and J.~Wenninger in {\em {Proceedings,
  5th International Particle Accelerator Conference (IPAC 2014): Dresden,
  Germany, June 15-20, 2014}}, p.~MOXAA01, 2014.

\bibitem{Kovchegov:2012mbw}
Y.~V. Kovchegov and E.~Levin, {\em {Quantum chromodynamics at high energy}}.
\newblock Cambridge University Press, 2012.

\bibitem{Aktas:2005xu}
{\bf H1} collaboration, A.~Aktas {\em et.~al.}, {\it {Elastic $J/\Psi$
  production at HERA}},
  \href{http://dx.doi.org/10.1140/epjc/s2006-02519-5}{{\em Eur. Phys. J.} {\bf
  C46} (2006) 585} [\href{http://arXiv.org/abs/hep-ex/0510016}{{\tt
  arXiv:hep-ex/0510016}}].

\bibitem{Chekanov:2002xi}
{\bf ZEUS} collaboration, S.~Chekanov {\em et.~al.}, {\it {Exclusive
  photoproduction of $J/\Psi$ mesons at HERA}},
  \href{http://dx.doi.org/10.1007/s10052-002-0953-7}{{\em Eur. Phys. J.} {\bf
  C24} (2002) 345} [\href{http://arXiv.org/abs/hep-ex/0201043}{{\tt
  arXiv:hep-ex/0201043 [hep-ex]}}].

\bibitem{Berger:2011ew}
J.~Berger and A.~M. Stasto, {\it {Small x nonlinear evolution with impact
  parameter and the structure function data}},
  \href{http://dx.doi.org/10.1103/PhysRevD.84.094022}{{\em Phys. Rev.} {\bf
  D84} (2011) 094022} [\href{http://arXiv.org/abs/1106.5740}{{\tt
  arXiv:1106.5740 [hep-ph]}}].

\bibitem{Berger:2012wx}
J.~Berger and A.~M. Stasto, {\it {Exclusive vector meson production and small-x
  evolution}},  \href{http://dx.doi.org/10.1007/JHEP01(2013)001}{{\em JHEP}
  {\bf 01} (2013) 001} [\href{http://arXiv.org/abs/1205.2037}{{\tt
  arXiv:1205.2037 [hep-ph]}}].

\bibitem{Schlichting:2014ipa}
S.~Schlichting and B.~Schenke, {\it {The shape of the proton at high
  energies}},  \href{http://dx.doi.org/10.1016/j.physletb.2014.10.068}{{\em
  Phys. Lett.} {\bf B739} (2014) 313}
  [\href{http://arXiv.org/abs/1407.8458}{{\tt arXiv:1407.8458 [hep-ph]}}].

\bibitem{Kovchegov:1999yj}
Y.~V. Kovchegov, {\it {Small-$x$ $F_2$ structure function of a nucleus
  including multiple pomeron exchanges}},
  \href{http://dx.doi.org/10.1103/PhysRevD.60.034008}{{\em Phys. Rev.} {\bf
  D60} (1999) 034008} [\href{http://arXiv.org/abs/hep-ph/9901281}{{\tt
  arXiv:hep-ph/9901281 [hep-ph]}}].

\bibitem{Balitsky:1995ub}
I.~Balitsky, {\it {Operator expansion for high-energy scattering}},
  \href{http://dx.doi.org/10.1016/0550-3213(95)00638-9}{{\em Nucl. Phys.} {\bf
  B463} (1996) 99} [\href{http://arXiv.org/abs/hep-ph/9509348}{{\tt
  arXiv:hep-ph/9509348}}].

\bibitem{Albacete:2010sy}
J.~L. Albacete, N.~Armesto, J.~G. Milhano, P.~Quiroga-Arias and C.~A. Salgado,
  {\it {AAMQS: A non-linear QCD analysis of new HERA data at small-x including
  heavy quarks}},  \href{http://dx.doi.org/10.1140/epjc/s10052-011-1705-3}{{\em
  Eur. Phys. J.} {\bf C71} (2011) 1705}
  [\href{http://arXiv.org/abs/1012.4408}{{\tt arXiv:1012.4408 [hep-ph]}}].

\bibitem{Lappi:2013zma}
T.~Lappi and H.~M{\"a}ntysaari, {\it {Single inclusive particle production at
  high energy from HERA data to proton-nucleus collisions}},
  \href{http://dx.doi.org/10.1103/PhysRevD.88.114020}{{\em Phys. Rev.} {\bf
  D88} (2013) 114020} [\href{http://arXiv.org/abs/1309.6963}{{\tt
  arXiv:1309.6963 [hep-ph]}}].

\bibitem{Balitsky:2008zza}
I.~Balitsky and G.~A. Chirilli, {\it {Next-to-leading order evolution of color
  dipoles}},  \href{http://dx.doi.org/10.1103/PhysRevD.77.014019}{{\em Phys.
  Rev.} {\bf D77} (2008) 014019} [\href{http://arXiv.org/abs/0710.4330}{{\tt
  arXiv:0710.4330 [hep-ph]}}].

\bibitem{Lappi:2015fma}
T.~Lappi and H.~Mäntysaari, {\it {Direct numerical solution of the coordinate
  space Balitsky-Kovchegov equation at next to leading order}},
  \href{http://dx.doi.org/10.1103/PhysRevD.91.074016}{{\em Phys. Rev.} {\bf
  D91} (2015) 074016} [\href{http://arXiv.org/abs/1502.02400}{{\tt
  arXiv:1502.02400 [hep-ph]}}].

\bibitem{Lappi:2016fmu}
T.~Lappi and H.~Mäntysaari, {\it {Next-to-leading order Balitsky-Kovchegov
  equation with resummation}},
  \href{http://dx.doi.org/10.1103/PhysRevD.93.094004}{{\em Phys. Rev.} {\bf
  D93} (2016) 094004} [\href{http://arXiv.org/abs/1601.06598}{{\tt
  arXiv:1601.06598 [hep-ph]}}].

\bibitem{Lappi:2016oup}
T.~Lappi and R.~Paatelainen, {\it {The one loop gluon emission light cone wave
  function}},  \href{http://dx.doi.org/10.1016/j.aop.2017.02.002}{{\em Annals
  Phys.} {\bf 379} (2017) 34} [\href{http://arXiv.org/abs/1611.00497}{{\tt
  arXiv:1611.00497 [hep-ph]}}].

\bibitem{Boussarie:2016bkq}
R.~Boussarie, A.~V. Grabovsky, D.~{\relax Yu}. Ivanov, L.~Szymanowski and
  S.~Wallon, {\it {Next-to-Leading Order Computation of Exclusive Diffractive
  Light Vector Meson Production in a Saturation Framework}},
  \href{http://dx.doi.org/10.1103/PhysRevLett.119.072002}{{\em Phys. Rev.
  Lett.} {\bf 119} (2017) 072002} [\href{http://arXiv.org/abs/1612.08026}{{\tt
  arXiv:1612.08026 [hep-ph]}}].

\bibitem{Beuf:2017bpd}
G.~Beuf, {\it {Dipole factorization for DIS at NLO: Combining the $q\bar{q}$
  and $q\bar{q}g$ contributions}},
  \href{http://dx.doi.org/10.1103/PhysRevD.96.074033}{{\em Phys. Rev.} {\bf
  D96} (2017) 074033} [\href{http://arXiv.org/abs/1708.06557}{{\tt
  arXiv:1708.06557 [hep-ph]}}].

\bibitem{Ducloue:2017ftk}
B.~Ducloué, H.~Hänninen, T.~Lappi and Y.~Zhu, {\it {Deep inelastic scattering
  in the dipole picture at next-to-leading order}},
  \href{http://dx.doi.org/10.1103/PhysRevD.96.094017}{{\em Phys. Rev.} {\bf
  D96} (2017) 094017} [\href{http://arXiv.org/abs/1708.07328}{{\tt
  arXiv:1708.07328 [hep-ph]}}].

\bibitem{Hanninen:2017ddy}
H.~Hänninen, T.~Lappi and R.~Paatelainen, {\it {One-loop corrections to light
  cone wave functions: the dipole picture DIS cross section}},
  \href{http://dx.doi.org/10.1016/j.aop.2018.04.015}{{\em Annals Phys.} {\bf
  393} (2018) 358} [\href{http://arXiv.org/abs/1711.08207}{{\tt
  arXiv:1711.08207 [hep-ph]}}].

\bibitem{Ball:2017nwa}
{\bf NNPDF} collaboration, R.~D. Ball {\em et.~al.}, {\it {Parton distributions
  from high-precision collider data}},
  \href{http://dx.doi.org/10.1140/epjc/s10052-017-5199-5}{{\em Eur. Phys. J.}
  {\bf C77} (2017) 663} [\href{http://arXiv.org/abs/1706.00428}{{\tt
  arXiv:1706.00428 [hep-ph]}}].

\bibitem{Aaron:2008ad}
{\bf H1} collaboration, F.~Aaron {\em et.~al.}, {\it {Measurement of the Proton
  Structure Function $F_L(x, Q^2)$ at Low $x$}},
  \href{http://dx.doi.org/10.1016/j.physletb.2008.05.070}{{\em Phys. Lett.}
  {\bf B665} (2008) 139} [\href{http://arXiv.org/abs/0805.2809}{{\tt
  arXiv:0805.2809 [hep-ex]}}].

\bibitem{Chekanov:2009na}
{\bf ZEUS} collaboration, S.~Chekanov {\em et.~al.}, {\it {Measurement of the
  Longitudinal Proton Structure Function at HERA}},
  \href{http://dx.doi.org/10.1016/j.physletb.2009.10.050}{{\em Phys. Lett.}
  {\bf B682} (2009) 8} [\href{http://arXiv.org/abs/0904.1092}{{\tt
  arXiv:0904.1092 [hep-ex]}}].

\bibitem{Ryskin:1992ui}
M.~Ryskin, {\it {Diffractive $J/\Psi$ electroproduction in LLA QCD}},
  \href{http://dx.doi.org/10.1007/BF01555742}{{\em Z. Phys.} {\bf C57} (1993)
  89}.

\bibitem{Bartels:2003yj}
J.~Bartels, K.~J. Golec-Biernat and K.~Peters, {\it {On the dipole picture in
  the nonforward direction}},  {\em Acta Phys. Polon.} {\bf B34} (2003) 3051
  [\href{http://arXiv.org/abs/hep-ph/0301192}{{\tt arXiv:hep-ph/0301192
  [hep-ph]}}].

\bibitem{Shuvaev:1999ce}
A.~G. Shuvaev, K.~J. Golec-Biernat, A.~D. Martin and M.~G. Ryskin, {\it
  {Off-diagonal distributions fixed by diagonal partons at small $x$ and
  $\xi$}},  \href{http://dx.doi.org/10.1103/PhysRevD.60.014015}{{\em Phys.
  Rev.} {\bf D60} (1999) 014015}
  [\href{http://arXiv.org/abs/hep-ph/9902410}{{\tt arXiv:hep-ph/9902410}}].

\bibitem{Lappi:2010dd}
T.~Lappi and H.~M{\"a}ntysaari, {\it {Incoherent diffractive
  $J/\Psi$-production in high energy nuclear DIS}},
  \href{http://dx.doi.org/10.1103/PhysRevC.83.065202}{{\em Phys. Rev.} {\bf
  C83} (2011) 065202} [\href{http://arXiv.org/abs/1011.1988}{{\tt
  arXiv:1011.1988 [hep-ph]}}].

\bibitem{Alexa:2013xxa}
{\bf H1} collaboration, C.~Alexa {\em et.~al.}, {\it {Elastic and
  Proton-Dissociative Photoproduction of J$/\Psi$ Mesons at HERA}},
  \href{http://dx.doi.org/10.1140/epjc/s10052-013-2466-y}{{\em Eur. Phys. J.}
  {\bf C73} (2013)~no.~6 2466} [\href{http://arXiv.org/abs/1304.5162}{{\tt
  arXiv:1304.5162 [hep-ex]}}].

\bibitem{Bertulani:2005ru}
C.~A. Bertulani, S.~R. Klein and J.~Nystrand, {\it {Physics of ultra-peripheral
  nuclear collisions}},
  \href{http://dx.doi.org/10.1146/annurev.nucl.55.090704.151526}{{\em Ann. Rev.
  Nucl. Part. Sci.} {\bf 55} (2005) 271}
  [\href{http://arXiv.org/abs/nucl-ex/0502005}{{\tt arXiv:nucl-ex/0502005
  [nucl-ex]}}].

\bibitem{Lappi:2013am}
T.~Lappi and H.~M{\"a}ntysaari, {\it {$J/\Psi$ production in ultraperipheral
  Pb+Pb and p+Pb collisions at LHC energies}},
  \href{http://dx.doi.org/10.1103/PhysRevC.87.032201}{{\em Phys. Rev.} {\bf
  C87} (2013) 032201} [\href{http://arXiv.org/abs/1301.4095}{{\tt
  arXiv:1301.4095 [hep-ph]}}].

\bibitem{Mantysaari:2016ykx}
H.~Mäntysaari and B.~Schenke, {\it {Evidence of strong proton shape
  fluctuations from incoherent diffraction}},
  \href{http://dx.doi.org/10.1103/PhysRevLett.117.052301}{{\em Phys. Rev.
  Lett.} {\bf 117} (2016) 052301} [\href{http://arXiv.org/abs/1603.04349}{{\tt
  arXiv:1603.04349 [hep-ph]}}].

\bibitem{Mantysaari:2016jaz}
H.~Mäntysaari and B.~Schenke, {\it {Revealing proton shape fluctuations with
  incoherent diffraction at high energy}},
  \href{http://dx.doi.org/10.1103/PhysRevD.94.034042}{{\em Phys. Rev.} {\bf
  D94} (2016)~no.~3 034042} [\href{http://arXiv.org/abs/1607.01711}{{\tt
  arXiv:1607.01711 [hep-ph]}}].

\bibitem{Armesto:2014sma}
N.~Armesto and A.~H. Rezaeian, {\it {Exclusive vector meson production at high
  energies and gluon saturation}},
  \href{http://dx.doi.org/10.1103/PhysRevD.90.054003}{{\em Phys. Rev.} {\bf
  D90} (2014)~no.~5 054003} [\href{http://arXiv.org/abs/1402.4831}{{\tt
  arXiv:1402.4831 [hep-ph]}}].

\bibitem{TheALICE:2014dwa}
{\bf ALICE} collaboration, B.~B. Abelev {\em et.~al.}, {\it {Exclusive
  $\mathrm{J/}\psi$ photoproduction off protons in ultra-peripheral p-Pb
  collisions at $\sqrt{s_{\rm NN}}=5.02$ TeV}},
  \href{http://dx.doi.org/10.1103/PhysRevLett.113.232504}{{\em Phys. Rev.
  Lett.} {\bf 113} (2014) 232504} [\href{http://arXiv.org/abs/1406.7819}{{\tt
  arXiv:1406.7819 [nucl-ex]}}].

\bibitem{Aaron:2009xp}
{\bf H1} collaboration, F.~Aaron {\em et.~al.}, {\it {Diffractive
  Electroproduction of $\rho$ and $\phi$ Mesons at HERA}},
  \href{http://dx.doi.org/10.1007/JHEP05(2010)032}{{\em JHEP} {\bf 1005} (2010)
  032} [\href{http://arXiv.org/abs/0910.5831}{{\tt arXiv:0910.5831 [hep-ex]}}].

\bibitem{Amaudruz:1995tq}
{\bf New Muon} collaboration, P.~Amaudruz {\em et.~al.}, {\it {A Reevaluation
  of the nuclear structure function ratios for D, He, Li-6, C and Ca}},
  \href{http://dx.doi.org/10.1016/0550-3213(94)00023-9}{{\em Nucl. Phys.} {\bf
  B441} (1995) 3} [\href{http://arXiv.org/abs/hep-ph/9503291}{{\tt
  arXiv:hep-ph/9503291 [hep-ph]}}].

\bibitem{Adams:1995is}
{\bf E665} collaboration, M.~R. Adams {\em et.~al.}, {\it {Shadowing in
  inelastic scattering of muons on carbon, calcium and lead at low $x_{Bj}$}},
  \href{http://dx.doi.org/10.1007/BF01624583}{{\em Z. Phys.} {\bf C67} (1995)
  403} [\href{http://arXiv.org/abs/hep-ex/9505006}{{\tt arXiv:hep-ex/9505006
  [hep-ex]}}].

\bibitem{Arneodo:1995cs}
{\bf New Muon} collaboration, M.~Arneodo {\em et.~al.}, {\it {The Structure
  Function ratios $F_2(Li) / F_2(D)$ and $F_2(C) / F_2(D)$ at small $x$}},
  \href{http://dx.doi.org/10.1016/0550-3213(95)00023-2}{{\em Nucl. Phys.} {\bf
  B441} (1995) 12} [\href{http://arXiv.org/abs/hep-ex/9504002}{{\tt
  arXiv:hep-ex/9504002 [hep-ex]}}].

\bibitem{Kowalski:2007rw}
H.~Kowalski, T.~Lappi and R.~Venugopalan, {\it {Nuclear enhancement of
  universal dynamics of high parton densities}},
  \href{http://dx.doi.org/10.1103/PhysRevLett.100.022303}{{\em Phys. Rev.
  Lett.} {\bf 100} (2008) 022303} [\href{http://arXiv.org/abs/0705.3047}{{\tt
  arXiv:0705.3047 [hep-ph]}}].

\bibitem{Mantysaari:2017slo}
H.~Mäntysaari and R.~Venugopalan, {\it {Systematics of strong nuclear
  amplification of gluon saturation from exclusive vector meson production in
  high energy electron–nucleus collisions}},
  \href{http://dx.doi.org/10.1016/j.physletb.2018.04.044}{{\em Phys. Lett.}
  {\bf B781} (2018) 664} [\href{http://arXiv.org/abs/1712.02508}{{\tt
  arXiv:1712.02508 [nucl-th]}}].

\end{thebibliography}\endgroup

\end{document}